\documentclass[a4paper,11pt]{article}
\usepackage{jcappub}

 \usepackage[utf8]{inputenc}

\usepackage{mathtools}
\usepackage{booktabs}
\usepackage[english]{babel}
\usepackage{amsmath,amssymb,amsbsy,amstext, amsthm, simplewick, amsfonts}
\usepackage{hyperref}
\usepackage{graphicx}
\usepackage[small]{caption}
\usepackage{siunitx}
\usepackage{upgreek}
\usepackage{framed}
\usepackage{wrapfig}
\usepackage{multirow}
\usepackage{bbm}
\usepackage[svgnames,dvipsnames,x11names]{xcolor}

\usepackage{selinput}

\usepackage{bm}
\usepackage{float}
\usepackage[top=2.5cm,bottom=3.cm,right=2.5cm,left=2.5cm]{geometry}
\usepackage{yfonts}
\usepackage{caption}
\usepackage{subcaption}
\usepackage{sidecap}
\usepackage{longtable}
\usepackage{anyfontsize}
\usepackage{dsfont}
\usepackage{tikz}
\usepackage{relsize}
\usepackage{tcolorbox}

\usepackage{xparse}

\usepackage{slashed}
\usepackage{simpler-wick}

\usepackage{shorthand}

\usepackage{array}
\newcolumntype{L}[1]{>{\raggedright\let\newline\\\arraybackslash\hspace{0pt}}m{#1}}
\newcolumntype{C}[1]{>{\centering\let\newline\\\arraybackslash\hspace{0pt}}m{#1}}
\newcolumntype{R}[1]{>{\raggedleft\let\newline\\\arraybackslash\hspace{0pt}}m{#1}}

\usepackage[framemethod=default]{mdframed}
\newmdenv[skipabove=7pt,
skipbelow=7pt,
rightline=false,
leftline=false,
topline=false,
bottomline=false,
backgroundcolor=gray!10,
linecolor=gray,
innerleftmargin=5pt,
innerrightmargin=5pt,
innertopmargin=5pt,
innerbottommargin=5pt,
leftmargin=0cm,
rightmargin=0cm,
linewidth=4pt]{eBox}
\newmdenv[skipabove=7pt,
skipbelow=7pt,
rightline=false,
leftline=false,
topline=false,
bottomline=false,
backgroundcolor=gray!10,
linecolor=gray,
innerleftmargin=5pt,
innerrightmargin=5pt,
innertopmargin=-5pt,
innerbottommargin=5pt,
leftmargin=0cm,
rightmargin=0cm,
linewidth=4pt]{eBox2}
\usepackage[framemethod=default]{mdframed}
\newmdenv[skipabove=7pt,
skipbelow=7pt,
rightline=true,
leftline=true,
topline=true,
bottomline=true,
backgroundcolor=gray!15,
linecolor=gray,
innerleftmargin=5pt,
innerrightmargin=5pt,
innertopmargin=5pt,
innerbottommargin=5pt,
leftmargin=0cm,
rightmargin=0cm,
linewidth=0.75pt]{eBox3}

\definecolor{lightgreen}{cmyk}{0.2, 0, 0.2, 0.2}
\definecolor{lightgray2}{cmyk}{0.1,0.1,0,0.1}
\definecolor{Red2}{RGB}{214, 39, 40}
\definecolor{Blue2}{RGB} {31, 119, 180}
\definecolor{Green2}{RGB}{44, 160, 44}

\newcommand{\Comment}[1]{{}}

\newcommand   \Kpv {\ensuremath{K_{\mathsmaller{\rm PV}}}}
\newcommand   \ddm {\ensuremath{\avg{\de\de p_\parallel}} }
\allowdisplaybreaks[1]
\usepackage{CJKutf8}

\title{Boosted Odds:\texorpdfstring{\\}{ } Lower-Point Probes of Parity Violation with Momentum Fields}
\author[ab]{Yong Sheng Yap \begin{CJK*}{UTF8}{bsmi}(葉勇陞)\end{CJK*},}
\author[cba]{William R. Coulton,}
\author[bde]{Jiamin Hou,}
\author[ab]{Steven Gratton}
\emailAdd{ysy23@cam.ac.uk}

\affiliation[a]{ DAMTP, Centre for Mathematical Sciences, University of Cambridge, Wilberforce Road, Cambridge CB3 0WA, UK.}
\affiliation[b]{Kavli Institute for Cosmology Cambridge, Madingley Road, Cambridge CB3 0HA, UK.}
\affiliation[c]{Department of Physics, University of Oxford, Denys Wilkinson Building, Keble Road, Oxford OX1 3RH, UK.}
\affiliation[d]{Faculty of Physics, Ludwig-Maximilians-Universit{\"a}t, Scheinerstr. 1, 81679 Munich, Germany.}
\affiliation[e]{Institute of Astronomy, University of Cambridge, Madingley Rd, Cambridge CB3 0HA, UK} 

\abstract{
  Parity-violating initial conditions are uniquely compelling signatures of inflationary dynamics beyond the single-field slow-roll paradigm, and their effects on the distribution of late-time observables can be easily separated from that of gravitational nonlinearities and galaxy formation physics. In this work, we propose using the momentum field, $\mathbf{p} = (1+\delta)\mathbf{v}$, and, in particular, the density-density-momentum bispectrum as a novel probe of primordial parity violation. 
  This class of bispectra greatly complements traditional searches using four-point statistics of the density field, furnishing new observables that can be sensitive to parity-violating physics. At the same time, its lower dimensionality substantially reduces the complexity associated with analysing high-dimensional data vectors and their covariance. We show that the sensitivity of the density-density-momentum bispectrum to parity violation effects persists even if we can only observe the projection of the momentum field along the line-of-sight. We perform a Fisher forecast for the parity-violating amplitude of the ultra-squeezed template in the cross bispectrum of kinetic Sunyaev-Zel'dovich (kSZ) anisotropies and galaxy clustering, and we find that our approach give competitive constraints when compared to existing methods. Our method can be easily co-opted to existing pipelines to study parity violation and, as the galaxy-galaxy-kSZ bispectrum is a key cosmological probe, take advantage of the significant existing efforts dedicated to mitigating systematics and covariance modelling.
  }

\begin{document}

\maketitle
\flushbottom

\section{Introduction}\label{sec:intro}

Parity is not a symmetry of our Universe today \cite{Lee:1956qn, Wu:1957my}. The violation of parity in the Standard Model leads to many intriguing features in the low-energy physics we observe. This then invites the question of whether parity was ever a symmetry at higher energy scales in the distant past, and, if not, the follow-up question of whether the Universe might provide a view of parity violation through cosmological observations today.

\vskip 5pt
The inflationary paradigm has provided us with the most promising explanation for the origin of structure, in which vacuum fluctuations are stretched to superhorizon sizes and seed the fluctuations in the large-scale distribution of matter. As such, the matter field inherits correlations generated by dynamics in the primordial Universe. Assuming that the late Universe follows the parity-invariant laws of gravity and hydrodynamics, ascertaining a violation of parity in its large-scale correlations would cast a direct spotlight on the microphysics that underpin inflation. This is especially interesting as theoretical studies \cite{Liu:2019fag,Cabass:2022rhr,Thavanesan:2025kyc} have found that observable levels of parity violation are not manifest in the vanilla single-field slow-roll paradigm, and a confirmed detection will be highly indicative of new forces or particles in play during the inflationary epoch \cite{Shiraishi:2016mok,Cabass:2021fnw,Tong:2022cdz,Niu:2022fki,Fujita:2023inz,Stefanyszyn:2025yhq,Cho:2025rvg}. Joint with the fact that parity-violating correlations are symmetry-protected, i.e.\ their signal cannot be mimicked by gravitational evolution, they constitute a clean channel of primordial non-Gaussianity.

\vskip 5pt
The primary obstacle to template-blind searches of parity violation is the high-dimensionality of its estimators. Current searches \cite{Hou:2022wfj,Philcox:2022hkh,Adari:2024vkf,Slepian:2025kbb,Hou:2025cey} rely on the four-point correlation function (4PCF) first proposed in \cite{Cahn:2021ltp} as the main statistic to diagnose parity violation, which suffers from substantial difficulties in accurately modelling its covariance \cite{Hou:2022wfj, Hou:2025cey}.  Nevertheless, mitigating techniques can been developed to overcome these challenges, cross-patch statistics have been introduced to reduce bias from mismatches between the data and mock catalogues \cite{Krolewski:2024paz}. Other ongoing pursuits in this direction have taken shape in terms of finding lower-dimensional estimators, such as parity-odd power spectra of composite fields \cite{Jeong:2012df,Jamieson:2024mau,Gao:2025yqd,Gao:2026zaz}, which are built of kernels tailored to optimally match signals of interest. Wise choice of kernels can effectively single out parity-violating contributions and allow for efficient searches. 

\vskip 5pt
For the reasons listed above, it is evident that we must be careful when evaluating claims of detection for parity violation. To robustly confirm or exclude parity violation, it is crucial that results are consistent across complementary probes of the initial conditions. At the level of 4-point statistics, recent work has been done to understand signatures of parity violation in the cosmic microwave background (CMB) and its secondary anisotropies \cite{Philcox:2023ffy,Philcox:2023ypl,Philcox:2025bvj,Philcox:2025wts,Greco:2025xtt,Slepian:2025ykz}. In conjunction, systematic studies have also been conducted for the effects of parity violation on the angular momenta of virialised structures \cite{Coulton:2023oug,Shim:2024tue,daSilveiraFerreira:2026fdy},
and in the intrinsic alignment and lensing of galaxy shapes \cite{Kurita:2025hmp,Chen:2026wok} via analytical methods, both of which are rising in importance as independent probes of cosmology as the data improves. Complementarily, increasing resolution in CMB experiments and galaxy surveys has enabled better sampling of the peculiar velocity field independently of its effect on the observed galaxy distributions. Many efforts have been devoted to extract velocity information out of a plethora of tracers, ranging from magnitude fluctuations of standard candles \cite{Hui:2005nm,Johnson:2014kaa,Macaulay:2016uwy,Huterer:2016uyq,Stahl:2021mat,Do:2024iuw,Nguyen:2025gfc}, redshift-independent distance indicators like the Tully-Fisher \cite{Tully:1977fu,Kourkchi:2022ifq,Douglass:2025jnn} and Fundamental Plane relations \cite{Djorgovski:1987vx,Dressler:1987ny,Campbell:2014uia,Springob:2014qja,Howlett:2022len,Ross:2025lzj}, to secondary anisotropies in the CMB \cite{Sunyaev:1972eq,Sunyaev:1980vz, Birkinshaw:1983,Ostriker:1986fua, Gurvits1986}. Despite these observational advances, the influence of parity-violating initial conditions (PVICs) on velocity fields remains an unexplored facet of this frontier. Motivated by this, we find it compelling to assess the sensitivity of velocity fields as a complementary probe of parity violation. 

\vskip 5pt
Before we start our analysis, we find it important to address fundamental limitations of velocity measurements. As our telescopes are limited by resolution and exposure time, measurements of the velocity field are inherently undersampled in low-density regions in our Universe. Rather than attempting to correct for this, we consider instead the clustering-modulated velocity field $\bp \equiv (1+\de)\bv$ (the ``momentum'' field) as our natural observable throughout our work. Reflecting current progress on measuring radial velocities, we focus our attention to cases where only the line-of-sight (LoS) component of the velocity field is observed. Typically, as in the case for galaxy shapes \cite{Vlah:2019byq,Kurita:2025hmp}, non-scalar quantities are sensitive to parity-violating effects already at the level of the power spectrum. As such, partial access to a vectorial quantity also poses an intriguing dilemma to the sensitivity of lower-point velocity statistics to PVICs. Interestingly, within these constraints, we find that the bispectrum obtained by correlating the LoS momentum field with two density fields is parity-sensitive and this quantity will be the main focus of this work. Excitingly, this bispectrum is also a key component in existing CMB pipelines to extract correlations between the kinetic Sunyaev-Zel'dovich anisotropies and large-scale galaxy clustering \cite{SimonsObservatory:2018koc,Chaves-Montero:2019isa,Tanimura:2020une,Contreras:2022zdz,Bloch-2024,Krywonos:2024mpb,Lague:2024czc,Hotinli:2025tul,Lai:2025qdw,McCarthy:2025brx,Chaussidon:2026vmn,Qu:2026zyh}. Hence, our work also presents a path to parity violation searches that can exploit the extensive work in modelling and systematics in upcoming CMB analyses. 

\vskip 5pt
This paper is organised as follows: We begin with a pedagogical introduction to parity violation and how symmetry informs the construction of parity-sensitive estimators in Section \ref{sec:pvics}. In Section \ref{sec:B_ddp}, we discuss our main observable, the density-density-momentum bispectrum, and the signature of parity-violating signals in its different forms. A Fisher forecast is done in Section \ref{sec:forecast}, with the kinetic Sunyaev-Zel'dovich effect as a probe of the momentum field. We compare our method to other probes of parity violation in Section \ref{sec:discuss} and conclude in Section \ref{sec:conclude}. Some mathematical and technical details are relegated to the appendices. Appendix \ref{sec:wig} compiles mathematical properties of Wigner symbols and Wigner $d$-matrices used in this work. In appendix \ref{sec:deriv}, we show that effects of parity-violating initial conditions do not enter the density-density-velocity bispectrum at all orders in Standard Perturbation Theory. We analytically compute the leading contributions to parity-violating initial conditions to the density-density-momentum bispectrum in toy models of scaling universes in appendix \ref{sec:scale}, and show that it does not trivially vanish after projection and angular averaging in appendix \ref{sec:ang-avg}. Finally, we detail our numerical implementation of the Fisher forecast in appendix \ref{sec:effeval}.

\subsection*{Notation and Conventions}
For brevity, we sometimes denote integrals by subscripting the variable integrated over, e.g.\ $\int\dd z~ f(z) \equiv \int_z f(z)$. For integrals over multiple momenta, we write
\begin{align}
    \int \frac{\dd^3\bq_1}{(2\pi)^3}\int \frac{\dd^3\bq_2}{(2\pi)^3} \cdots \int \frac{\dd^3\bq_n}{(2\pi)^3} ~f(\bq_1,\bq_2,\ldots,\bq_n)\equiv \int_{\bq_1,\bq_2,\ldots,\bq_n}\hspace{-3em}f(\bq_1,\bq_2,\ldots,\bq_n)\equiv \int_{\{\bq_a\}}f(\{\bq_a\})\;,
\end{align}
where we typically package functions and integrals over multiple similar variables as a set, as in the last definition. Angular integrals over position-space variables (e.g.\ the line-of-sight) are written as $\int\dd^2\nhat \equiv \int_{\nhat}$, and those over Fourier-space variables as $\int \dd^2\khat \equiv \int_{\khat}$.

\vskip 5pt
The sum of momenta is abbreviated as $\bk_{1\ldots n} \equiv \bk_1+\ldots+\bk_n$, and similarly for other quantities, e.g.\ their harmonic counterparts $\ell_{1\ldots n}\equiv \ell_1+\ldots+\ell_n$. $n$-point correlators stripped of the momentum-conserving Dirac delta function $(\de^D)$ are denoted with a prime, e.g.
\begin{align}
    \avg{\cO(\bk_1)\ldots\cO(\bk_n)} = (2\pi)^3\de^D(\bk_{1\ldots n}) \avg{\cO(\bk_1)\ldots\cO(\bk_n)}'\;.
\end{align}

\vskip 5pt
\paragraph{Nomenclature. }Rather inconveniently, the Fourier modes $\bk$ and the clustering modulated velocity field $\bp = (1+\de)\bv$ are both referred to as ``momentum'' in large-scale structure (LSS) literature. To distinguish the two, we exclusively refer to the latter as the \textit{momentum field}, and the bispectrum \ddm (later defined in \eqref{eq:ddp-pv}) as the \textit{density-density-momentum bispectrum} to avoid confusion. All other occurrences of the term ``momentum'' refer to the Fourier modes.

\section{Observables for parity-violating initial conditions}\label{sec:pvics}

We begin with a brief overview of inflationary physics that could give rise to parity violation signals in the scalar sector, and discuss its imprints on the late-time observables typically considered in current analyses. We put particular emphasis on the kinematic constraints enforced by the interplay of rotations and parity inversions, which directly inform the sensitivity of various correlation functions.

\subsection{Parity violation from inflation}\label{sec:PV-from-inf}
If parity-violating physics were present during inflation, its signature would compartmentalise into two distinct observables. Either they are encoded in the tensor fluctuations \cite{Soda:2011am,Bordin:2020eui,Cabass:2021fnw,Orlando:2022rih,Tong:2022cdz}, or they are imprinted through interactions onto the scalar sector in curvature or isocurvature perturbations. Despite the interesting phenomenology present in the tensor sector, as primordial gravitational waves remain undetected in the CMB, we will focus our attention on the scalar sector. In particular, we will further assume adiabatic initial conditions and only consider statistics of the curvature perturbations.

\vskip 5pt
Parity violation presents a compelling subclass of primordial non-Gaussianity to look for. Firstly, it falls under the category of ``protected observables'' \cite{Green:2022hhj}, i.e.\ features that are stable against small-scale nonlinearities and difficult to replicate in the late Universe.\footnote{We assume that structure formation is governed by General Relativity and hydrodynamics, which is parity-symmetric. Understandably, parity violation not attributed to inflation can be seen on cosmological scales if the spacetime-filling media (e.g.\ dark matter, dark energy) of the Universe violates parity \cite{Komatsu:2022nvu,Salcedo:2024nex,Salcedo:2025ezu}.} Furthermore, any detection (if attributed to inflation) would exclude all statistically homogeneous and isotropic single-field slow-roll models with a Bunch-Davies vacuum. This is a result of various no-go theorems \cite{Liu:2019fag,Cabass:2022rhr,Thavanesan:2025kyc} that prove that parity-violating interactions do not affect the perturbative statistics of curvature perturbations in such cosmologies. As such, known models of inflation that can exhibit observable levels of parity violation typically host extra degrees of freedom \cite{Arkani-Hamed:2015bza,Jazayeri:2023kji,Tong:2022cdz,Stefanyszyn:2023qov,Stefanyszyn:2025yhq}, e.g.\ higher-spin fields in axion-gauge inflation \cite{Niu:2022fki,Fujita:2023inz,Cho:2025rvg}; exhibit strong departure from slow-roll, e.g.\ through non-canonical kinetic terms in ghost inflation \cite{Arkani-Hamed:2003juy,Cabass:2022oap}; or are accompanied by the breakdown of isotropy \cite{Shiraishi:2016mok}. In addition, Regardless of the underlying mechanism, any detection would provide a profound insight extending knowledge about the state and dynamics during the inflationary epoch.

\vskip 5pt
\paragraph{Kinematic constraints.} Parity violation is a statistical asymmetry between a process and its parity-inverted counterpart. Supposing these effects were weak during inflation, this amounts to probing a non-vanishing contribution to the parity-odd (denoted ``PO'' below) part of $n$-point statistics of the curvature perturbations, i.e.\ in Fourier space
\begin{align}
    \avg{\z^n}_\mathsmaller{\rm PO} &\equiv\avg{\z(\bk_1)\z(\bk_2)\ldots\z(\bk_n)} - \avg{\z(-\bk_1)\z(-\bk_2)\ldots\z(-\bk_n)}\;.
\end{align}
Also since $\z$ is a real field and hence satisfies $\z(-\bk) = \z^*(\bk)$, we stress that the parity-odd part of $\z$ correlations in Fourier space are purely imaginary.
However, not all correlations are parity-sensitive. Assuming statistical homogeneity, polyspectra ($n$-point correlation functions in Fourier space) of comoving curvature perturbations must respect momentum conservation, i.e.\ 
\begin{align}
    \avg{\z(\bk_1)\z(\bk_2)\ldots\z(\bk_n)} = (2\pi)^3\de^D(\bk_{12\ldots n})B_n(\bk_1,\bk_2,\ldots,\bk_n)\;.
\end{align}
An immediate consequence is that there are only $n-1$ independent external momenta $\bk_i \in \R^3$ for a $n$-point statistic. If these momenta lie on a 2D plane, the reflected configuration $\{-\bk\}$ can always be related to the original configuration $\{\bk\}$ by a proper rotation about the axis perpendicular to the plane. Equivalently, this means momenta configurations which satisfy 
\begin{align}
    B_n(\{-\bk\}) = B_n(\{\mathsf{R}\, \bk\})
\end{align}
for a rotation matrix $\mathsf{R}\in SO(3)$ would yield a vanishing $\avg{\z^n}_\mathsmaller{\rm PO}$, as $B_n(\{\mathsf{R}\, \bk\})= B_n(\{\bk\})$ by statistical isotropy. Additionally, since at most $n-1$ momenta are independent, this enforces that the power spectrum and bispectrum of curvature perturbations are insensitive to parity violation.\footnote{Note that the momentum-conserving delta function and Bose symmetry in $\z$ is sufficient to guarantee that $\avg{\z^2}_\mathsmaller{\rm PO}$ vanishes.} This further implies that late-time probes of parity violation must derive their sensitivity from the parity-odd 4-point (or higher-point) functions of curvature perturbations.

\paragraph{The parity-violating trispectrum.} We will only consider the parity-violating signals that are in the 4-point statistics. In the perturbative regime, parity violation can be characterised as a departure from Gaussianity in the curvature perturbations which we parametrise as \footnote{We adopt the factor of 9/25 to make contact with conventional {\fnl}  definitions, which are defined in terms of the primordial Bardeen potential $\Phi$. This is the same normalisation used in the \texttt{Quijote-ODD} simulations \cite{Coulton:2023oug}. At superhorizon scales, the two are related by $\z = 5\Phi/3$.}
\begin{align}\label{eq:def-ng}
   \z(\bk) =  \z_G(\bk) + \frac{9}{25}\pnl\int_{\bq_1,\bq_2,\bq_3} (2\pi)^3\de^D(\bk-\bq_{123})\Kpv(\bq_1,\bq_2,\bq_3)\z_G(\bq_1)\z_G(\bq_2)\z_G(\bq_3)\;,
\end{align}
in which $\pnl$ is a real scalar which determines the amplitude of parity violation. The parity-violating (indicated by ``PV'' below) trispectrum is then 
\begin{align}\label{eq:pv-zeta}
    B^{\mathsmaller{\rm PV}}_4(\bk_1,\bk_2,\bk_3,\bk_4) = \frac{9}{25}\pnl \Kpv(\bk_1,\bk_2,\bk_3)P_\z(k_1)P_\z(k_2)P_\z(k_3)+{\rm  perms.}\;,
\end{align}
where $P_\z(k) \equiv B_2(\bk,-\bk) $ is the power spectrum of $\z$, with
$P_\z(k)\approx 2\pi^2A_sk^{-3}$  for approximate scale-invariance. The permutations in \eqref{eq:pv-zeta} are cyclic over all momenta $\bk_{1,2,3,4}$. The parity-violating kernel $\Kpv$ is model-dependent, but can be generically written in terms of the combination of a shape function that transforms trivially under parity and a scalar triple product. Explicitly, 
\begin{align}\label{eq:Kpv}
    \Kpv(\bk_1,\bk_2,\bk_3) = \ci \bk_1\vdot(\bk_2\times\bk_3) S(k_1,k_2,k_3,\mu_{12},\mu_{23},\mu_{31})\;,
\end{align}
where the shape function $S$ depends only on the magnitude of the momenta, and their relative orientation $\mu_{ab} = \khat_a\vdot\khat_b$ to conform with rotational invariance. This reinforces the previous discussion of kinematic constraints, that the parity-odd information is encoded in the non-planar structure of the trispectrum. If parity-violating signals are isolated from other forms of non-Gaussianities, optimal estimators can be constructed to constrain the amplitudes of various parity-violating shapes. However, unlike in searches for primordial non-Gaussianity where templates have been designed to capture generic features from broad classes of inflationary physics \cite{Komatsu:2001rj,Babich:2004gb,Creminelli:2005hu,Senatore:2009gt,Suman:2025vuf,Suman:2025tpv}, exploration of the parity-violating landscape remain in its infancy. Hence for the rest of the paper, we will not commit to a particular model until section \ref{sec:forecast} in which a forecast is performed. In the latter case, we employ the template \cite{Coulton:2023oug}
\begin{align}\label{eq:will-temp}
    S(k_1,k_2,k_3,\mu_{12},\mu_{23},\mu_{31}) = k_1^{n_1}k_2^{n_2}k_3^{n_3}\;.
\end{align}
Although phenomenological in origin, this template has the benefits of being separable, and also scale-invariant if the indices satisfy $\sum_i n_i = -3$. Subject to this constraint, we have chosen the indices to take values $(n_1,n_2,n_3) = (-2,-1,0)$. This leads to a template that peaks in the ultra-squeezed regime (faster than local), which is reminiscent of the tachyonic collider \cite{McCulloch:2024hiz} and can be viewed as an archetype for models that exhibit large instabilities such as ghost inflation \cite{Arkani-Hamed:2003juy, Cabass:2022oap}. 

\subsection{Late-time observables}

Searches for parity violation make use of statistics capable of isolating the parity-violating trispectrum \eqref{eq:pv-zeta}. Primarily, they search for connected 4-point correlations in the fluctuations in matter and radiation. On large scales, scalar observables, such as overdensities, are linearly related to the curvature perturbations by a transfer function,
\begin{align}\label{eq:d=Mz}
   \de(\bk,z) = \cM_\de(k,z) \z(\bk)\;.
\end{align}
As such, for the clustering of biased tracers and CMB anisotropies, their correlations directly inherit the kinematic constraints of curvature perturbations on sufficiently large scales. Hence, the leading statistic for parity-violating signals of scalar observables also lie in the imaginary part of the trispectrum. This has been the main property exploited by model-agnostic approaches in many parity violation searches \cite{Hou:2022wfj,Philcox:2022hkh,Philcox:2023ffy,Philcox:2023ypl,Hewson:2024rnb,Adari:2024vkf,Slepian:2025kbb,Hou:2025cey}.

\paragraph{Composite Fields.} A complementary perspective is to look at 2-point correlations of composite fields. These are weighted products of the original scalar observable $\de$ evaluated at coincident points, such as $\de^2(\bx,z)$. In Fourier space, a general rank-$m$ composite field can be obtained via a convolution
\begin{align}\label{eq:CF-def}
    C^{i_1i_2\ldots i_m}_n[\de_\L](\bk,z) = \int_{\bq_1,\bq_2,\ldots \bq_n}\hspace{-3em}(2\pi)^3\de^D(\bk-\bq_{12\ldots n}) K^{i_1i_2\ldots i_m}_n(\bq_1,\bq_2,\ldots,\bq_n)\lr[{\prod_{a=1}^n \de_\L(\bq_a,z) }]\;,
\end{align}
where $K^{i_1i_2\ldots i_m}_n(\{\bq\})$ is a kernel constructed from the set of momenta engineered to provide sensitivity to the desired signature. Note that observables entering the convolution are typically smoothed at a resolution $\L$ to retain only large-scale correlations where \eqref{eq:d=Mz} is valid. 
% This folds in the symmetries of the primordial correlations into correlation functions of the composite fields. 
In the following, we restrict ourselves to the simple case where kernels are simple products of momenta, i.e.\ 
\begin{align}
    K_n^{i_1i_2\cdots i_m} \propto \prod_{a=1}^n \prod_{\a = 1}^{m_a} q_a^{i_\a}\;,
\end{align}
as they make up the basis for any kernel analytic in the momenta. The total number of momenta in the kernel can in general be larger than the rank, i.e.\ $\sum_{a=1}^{n}{m_a} \geq m$, as long as they are contracted with rotation-invariant tensors, i.e.\ the Kronecker delta $\de_{ij}$ or the Levi-Civita symbol $\eps_{ijk}$.

\vskip 5pt
Composite field correlators derive their sensitivity to the parity-odd trispectrum \eqref{eq:Kpv} through the scalar triple product. Particular observables of interest are angle-averaged power spectra\footnote{Statistics of composite fields should be renormalised by appropriate counterterms. This will hold true as well for the momentum field statistics we will discuss in section \ref{sec:B_ddp} \cite{Raikman:2026gcx}. Since renormalisation generates \emph{all} operators that obey the symmetries, template-based searches could suffer from shape mismatches and have biased results if this is not accounted for. This is beyond the scope of this paper, and since renormalisation does not impact model-agnostic searches of parity violation, we will neglect renormalisation in this work.} \cite{Jamieson:2024mau,Gao:2025yqd,Gao:2026zaz}
\begin{align}\label{eq:comp-pk}
    \int \frac{\dd^2\khat}{4\pi}\int \frac{\dd^2\khat'}{4\pi}\avg{C^{i_1\ldots i_m}_{n_1}[\de_\L](\bk)C^{i_1\ldots i_m}_{n_2}[\de_\L](\bk')}
\end{align}
with $n_1+n_2=4$. This expression contains the angular integrals 
\begin{align}\label{eq:composite-comp}
    \int \prod_{a=1}^4 \lr[{ \frac{\dd^2\qhat_a}{(2\pi)^2} \prod_{\a=1}^{m_a} q^{i_\a}_a }] F(q_1,q_2,q_3,\mu_{12},\mu_{23},\mu_{31}) \times 
    \begin{cases}
        1, & \text{for }\avg{\z^4}_\mathsmaller{\rm PO} =0 \\
        \bq_1\vdot(\bq_2\times \bq_3), & \text{for } \avg{\z^4}_\mathsmaller{\rm PO}\neq0
    \end{cases}
    \;,
\end{align}
where $F$ is built from transfer functions and the primordial shape (which is the function $S$ \eqref{eq:Kpv} for $\avg{\z^4}_\mathsmaller{\rm PO} \neq 0$ case, and the normalised trispectrum for $\avg{\z^4}_\mathsmaller{\rm PO} =0$.). The integral vanishes identically for an odd number of momentum insertions. Since the scalar triple product in $\avg{\z^4}_\mathsmaller{\rm PO}$  supplies three momentum insertions, kernel combinations of $C_{n_1}$ and $C_{n_2}$ that also provide a total of odd number of momenta in \eqref{eq:composite-comp}, i.e.\ $\sum_{a=1}^{4} m_a = {\rm odd}$, can effectively capture parity-violating trispectra and null contributions from parity-symmetric components. To form a scalar statistic \eqref{eq:comp-pk}, the odd number of momenta require contractions with a Levi-Civita symbol, mirroring the statistics constructed in \cite{Jamieson:2024mau, Gao:2025yqd, Gao:2026zaz}, which we refer to the readers for more details.

\section{Density-density-momentum bispectra}\label{sec:B_ddp}

We now study velocity fields in cosmology, and argue that generic observables that are affected by peculiar motion can be sensitive to parity violation. 

\subsection{Momentum fields in cosmology}\label{sec:vel-fields}

\begin{figure}
        \centering
        \includegraphics[width=0.6\linewidth]{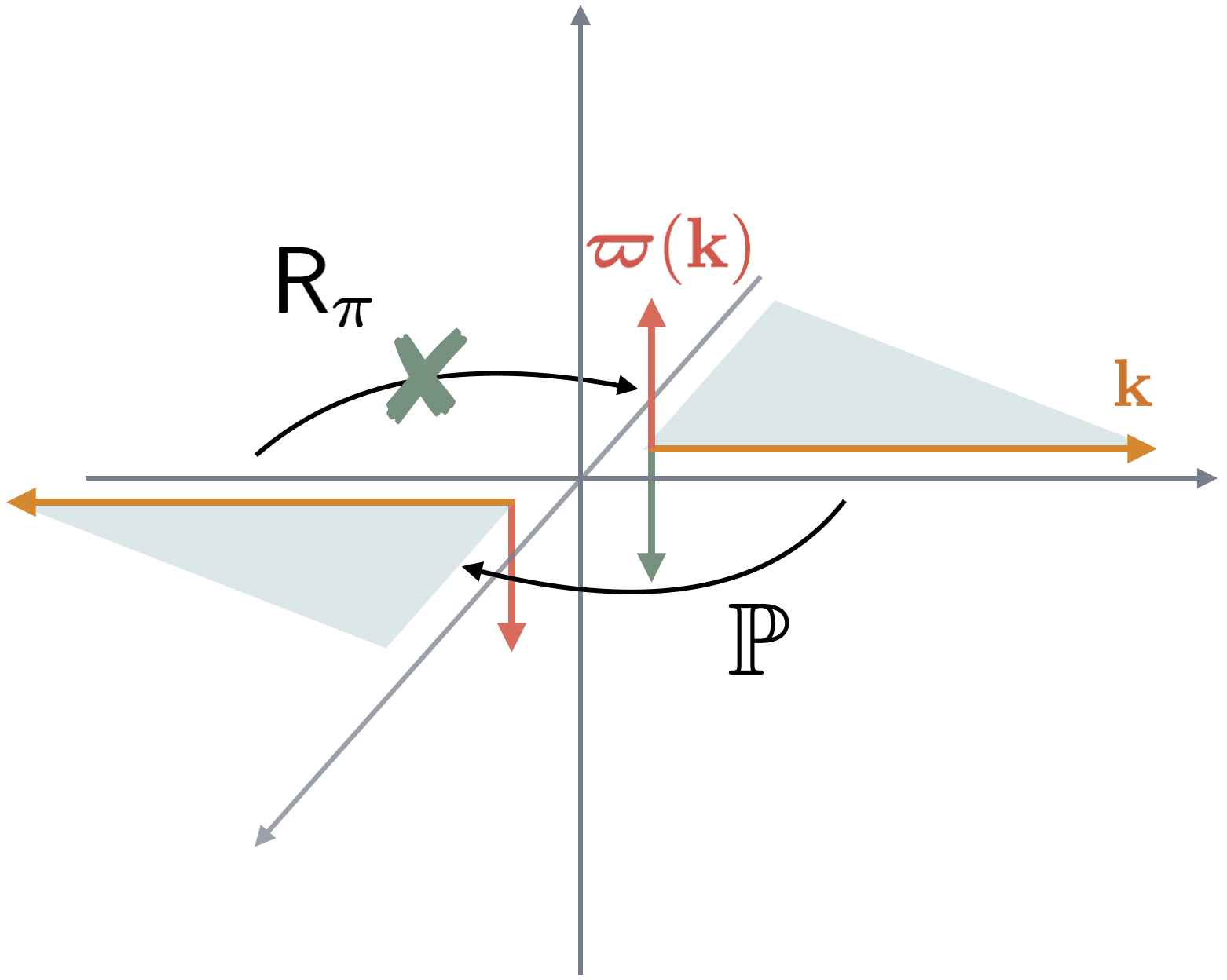}
        \caption{Schematic illustration of the parity-sensitivity of density-density-momentum bispectra. The triangle is formed by the momenta of the density-density-momentum bispectrum. The Fourier mode $\bk$ of the momentum field $\bp(\bk)$ is indicated in orange, and its transverse component $\boldsymbol{\varpi}(\bk)$, in red, is perpendicular to the plane of the triangle. The parity inverted configuration, given by the transformation $\mathbb{P}$, cannot be related to the original configuration via a proper rotation. For example, the $180^{\circ}$ rotation $\mathsf{R}_\pi$  in the plane of the triangle maps the parity inverted triangle back to the original triangle, but maps $\mathbb{P}[\boldsymbol{\varpi}(\bk)]$ to $-\boldsymbol{\varpi}(\bk)$, shown in green. }
        \label{fig:ddp}
    \end{figure}

In principle, the velocity of the matter field, which is related to the overdensity field via the continuity equation
\begin{align}\label{eq:cont-eq}
    \ptl_\eta\de = -  \nabla\vdot\lr[{\bv(1+\de)}]\;,
\end{align}
encodes the dynamical response of matter, and furnishes a complementary probe of large-scale structure. Nonetheless, the three-dimensional velocity field is practically impossible to measure directly, and can only be sparsely recovered at luminous sources near density peaks.\footnote{This statement is only applicable for direct measurements. Continuity equation velocity reconstruction and related inference methods  \cite{1991ApJ...372..380Y,Nusser:1993sx,Fisher:1994xm,Ho:2009iw,Shao:2010md,ACTPol:2015teu,Hernandez-Monteagudo:2019epd,Guachalla:2023lbx,Hadzhiyska:2023nig} do not fall under this umbrella.} As such, the meaningful observable is the clustering-modulated velocity, i.e.
\begin{align}\label{eq:momentum-def}
\bp(\bx,z) = \bv(\bx,z)[1+\de(\bx,z)]\;,
\end{align}
which the LSS community commonly refers to as the momentum field \cite{Seljak:2011tx,Okumura:2013zva,Pajer:2013jj,Sugiyama:2015dsa,Chen:2025aqr}. We propose the density-density-momentum bispectrum as a quantity that is sensitive to PVICs. The argument here is twofold:
\begin{enumerate}
    \item The velocity field we observe is nonlinearly modulated by the density field. Hence, even working with smoothed fields \cite{Baumann:2010tm}, the quadratic term $\bv\de$ in \eqref{eq:momentum-def} guarantees that the density-density-momentum bispectrum retains signals from the primordial trispectrum at one-loop. 
    \item In Fourier space, one could perform a Helmholtz decomposition of the momentum field 
    \begin{align}
        \bp(\bk) = \ci\bk \psi(\bk) + \boldsymbol{\varpi}(\bk)\;,
    \end{align}
    where $\boldsymbol{\varpi}(\bk)$ is divergence-free, i.e.\ $\bk \vdot\boldsymbol{\varpi}(\bk) = 0$. Unlike the velocity field, which can be largely regarded as irrotational at large scales due to the $a^{-1}$ decay of the vorticity, the same cannot be said about the momentum field. In fact, both the longitudinal and transverse components  contribute roughly equal amounts to the momentum field \cite{Mercolli:2013bsa}, largely due to the mode mixing inherent in the density modulation. In the density-density-momentum bispectrum where the Fourier modes are confined to a plane, $\boldsymbol{\varpi}(\bk)$ lies orthogonal to all the Fourier modes (See figure \ref{fig:ddp}). Hence, the density-density-momentum bispectrum is able to carry non-planar information from $\avg{\z^4}_\mathsmaller{\rm PO}$ via the extra degree of freedom provided by $\boldsymbol{\varpi}$. Note that the modulation on the velocity field is essential for sensitivity to PVICs: it can be shown explicitly that the parity-violating contribution to the density-density-velocity bispectrum vanishes identically at all loops in Standard Perturbation Theory (SPT), for which the derivation is left to appendix \ref{sec:deriv}.\footnote{Vorticity can be generated through the velocity dispersion of the dark matter fluid \cite{Pueblas:2008uv,Mercolli:2013bsa,Cusin:2016zvu,Jelic-Cizmek:2018gdp}, allowing the velocity field to also support a transverse component. This can be systematically accounted for in the framework of the Effective Field Theory of Large-Scale Structure \cite{Baumann:2010tm,Carrasco:2012cv,Hertzberg:2012qn,Anastasiou:2025jsy} (see \cite{Cabass:2022avo} for a recent review). However, since our focus is on {\ddm} which do have a non-vanishing transverse component, this does not impact any subsequent arguments, and we will stick to SPT for the rest of the paper.}
\end{enumerate}

In many observations, we do not observe all components of the momentum field. Typical methods measure only its projection along the line-of-sight (LoS) $\nhat$. Thus, it remains to show that the encoding of the non-planar structure in the density-density-momentum bispectrum is not lost when we have only partial access to its three-dimensional structure. We devote the rest of this section to this analysis.

\subsection{3D observables}\label{sec:plane-parallel}

To start with, we briefly describe 3D observables in real space. For simplicity, we consider the plane-parallel limit in which the region of the sky observed is sufficiently small that the variation of the LoS vector $\nhat$ across the sky can be neglected. 
% An important note is that the LoS in this approximation is a globally-defined entity, and \textit{does not transform} either under translations, rotations or parity inversions. 
Quantities projected onto the LoS will be denoted with a subscript $\parallel$, such as $p_\parallel \equiv \nhat\vdot\bp$.

\vskip 5pt
Hereinafter, we will refer to the density-density-momentum bispectrum projected onto the LoS, i.e.\ 
\begin{align}\label{eq:ddp}
    \avg{\de(\bk_1)\de(\bk_2)p_\parallel(\bK)} = \avg{\de(\bk_1)\de(\bk_2)v_\parallel(\bK)} + \avg{\de(\bk_1)\de(\bk_2)[\delta v_\parallel](\bK)}\;,
\end{align}
as \ddm for brevity. As aforementioned, the leading order contribution from the parity-violating primordial trispectra appears as a one-loop correction to the second term in \eqref{eq:ddp}, namely
\begin{align}
    \avg{\de(\bk_1)\de(\bk_2)p_\parallel(\bK)}_{\mathsmaller{\rm PV}}' = \ci f\cH  \int_{\bk_3} \frac{\nhat\vdot\khat_3}{k_3} \avg{\de(\bk_1)\de(\bk_2)\de(\bk_3)\de(\bk_4)}_\mathsmaller{\rm PV}\;,
\end{align}
where we write $\bK-\bk_3$ as $\bk_4$. Here, $f$ denotes the growth rate and $\cH$ the conformal Hubble parameter. At the linear level, $\avg{\de^4}_\mathsmaller{\rm PV }$ can be linearly related to $\avg{\z^4}_\mathsmaller{\rm PV }$ through use of the transfer functions \eqref{eq:d=Mz}. Using the parametrisation \eqref{eq:pv-zeta} and \eqref{eq:Kpv}, 
\begin{align}
    \avg{\de(\bk_1)\de(\bk_2)p_\parallel(\bK)}_{\mathsmaller{\rm PV}}' &= - \frac{9}{25} \pnl f\cH P_{1\z}(k_1)P_{1\z}(k_2)\nn\\
    &\hspace{1em}\times\int_{\bk_3} \frac{\nhat\vdot\khat_3}{k_3} \lr({\bk_1\vdot(\bk_2\times\bk_3)}) S(\{k_a,\mu_{ab}\}) P_{1\z}(k_3)\cM_\de(k_4)
\end{align}
plus cyclic permutations of $\bk_{1,2,3,4}$, and we have used the convention $P_{1\z}(k)\equiv \cM_{\de}(k) P_\z(k)$. Integrating over the internal momentum $\bk_3$, the parity-violating part of  \ddm is fixed by symmetry to take the form
\begin{align}\label{eq:ddp-pv}
    \avg{\de(\bk_1)\de(\bk_2) p_\parallel(\bK)}_{\mathsmaller{\rm PV}}' = \pnl (\nhat \vdot (\khat_1\times \khat_2) ) f\cH A_\mathsmaller{\rm PV}(k_1,k_2,K)\;,
\end{align}
with $A_\mathsmaller{\rm PV}(k_1,k_2,K)$ a model-dependent, real-valued function of the magnitude of the external momenta. For toy models of scale-free cosmologies \cite{Scoccimarro:1995if,Scoccimarro:1996se,Pajer:2013jj,Carrasco:2013sva,Carrasco:2013mua}, the scalar triple product in \eqref{eq:ddp-pv} (as well as the functional form of $A_\mathsmaller{\rm PV}$) can be computed analytically. We leave the details of the computation of these toy models to appendix \ref{sec:scale}. 

\vskip 5pt
The dependence of \eqref{eq:ddp-pv} on the pseudo-vector $\khat_1\times\khat_2$ is the distinctive signal of parity-violating contributions in large-scale correlations of the density and velocity fields, which is projected onto the LoS with the observables we measure. The scalar triple product is fixed by symmetry and therefore presents a common angular structure which all parity-violating signals in \ddm should embody. This universality also provides a foundation on which model-agnostic searches for parity violation in \ddm can be built. Moreover, the corresponding angular dependence of the bispectrum \eqref{eq:ddp-pv} can be naturally expressed in the Sugiyama \cite{Sugiyama:2018yzo} and Scoccimarro \cite{Scoccimarro:1999ed} bases for anisotropic bispectra. This suggests that existing frameworks do not require new innovations in estimator machinery and can be readily extended to search for parity-violating signals through \ddm in current and future synergies between galaxy and peculiar velocity surveys \cite{Courtois:2022hmr,Kourkchi:2022ifq,Do:2024iuw,Douglass:2025jnn,Ross:2025lzj,LSSTDarkEnergyScience:2025irx}.\footnote{To make for more efficient numerics, current implementations of bispectrum estimators do further impose parity-symmetric constraints to reduce computational cost. These assumptions would have to be relaxed when adapting the estimators for parity violation searches, but do not pose fundamental obstacles to searching for parity-violating contributions in \ddm observables.} 

\subsection{Projected observables}\label{sec:projected-obs}

We now go beyond the plane-parallel limit and consider observables on the full sky. There are many formalisms to extend plane-parallel results to the full sky \cite{1973ApJ...185..413P,Fisher:1994xm,Heavens:1994iq,Szalay:1997cc,Verde:2000xj,Wang:2020wsx,Benabou:2023ldb} that come with their respective strengths and weaknesses. In this work, we choose to work with projected fields and their corresponding angular bispectra. This choice is made as it has been the primary method used in cross-correlations with the CMB, whose secondary anisotropies can also be used to measure density and momentum fields in the late Universe.

\subsubsection{Symmetries on the sphere}\label{sec:sphere-symm}
We will focus on observables $\cO$, whose maps under rotations $\sf R$ on the sphere transform as
\begin{align}\label{eq:rot-inv}
    {\sf R}:~\cO(\nhat) \to \cO'(\nhat) = \cO({\sf R}^{-1}\nhat)\;.
\end{align}
This encompasses not only maps of CMB temperature, galaxy overdensities, but also maps of the LoS component of the momentum field.\footnote{Under rotations, a vector-valued function such as the momentum field $\bp$ transforms as \begin{align}
    {\sf R}:~\bp(\nhat) \to \bp'(\nhat) = {\sf R}\bp({\sf R}^{-1}\nhat)\;.
\end{align} Its projection onto the LoS, $p_\parallel(\nhat) \equiv \nhat\vdot\bp(\nhat)$ thus transforms to
\begin{align}
    p'_\parallel(\nhat) = \nhat\vdot \bp'(\nhat) = \nhat\vdot{\sf R}\bp({\sf R}^{-1}\nhat) = ({\sf R}^{-1}\nhat)\vdot\bp({\sf R}^{-1}\nhat) = p_\parallel({\sf R}^{-1}\nhat)
\end{align}
which satisfy the transformation property \eqref{eq:rot-inv}.
} For real-valued observables, their spherical harmonic decompositions have coefficients 
\begin{align}\label{eq:harm-coeff}
    \cO_{\ell m} = \int \dd^2\nhat~  Y^*_{\ell m}(\nhat)\cO(\nhat)\;,
\end{align}
satisfying $\cO^*_{\ell m}  = (-1)^m \cO_{\ell-m}$.
To respect statistical isotropy, the bispectrum should be invariant under rotations on the sphere. The angle-averaged bispectrum  $b_{\ell_1\ell_2\ell_3}$ is defined by \cite{Spergel:1999xn,Hu:2001fa}
\begin{align}\label{eq:red-bis-def}
    \avg{\cO_{\ell_1 m_1}\cO_{\ell_2 m_2}\cO_{\ell_3 m_3}} = \begin{pmatrix}
        \ell_1 & \ell_2 & \ell_3 \\
        m_1 & m_2 & m_3
    \end{pmatrix} b_{\ell_1\ell_2\ell_3}\;.
\end{align}
The quantity involving parentheses is a Wigner 3j symbol (see appendix \ref{sec:wig}), which enforces the triangle inequality amongst all the legs $\ell_i$ and a selection rule $\sum_{i=1}^{3}m_i = 0$. For brevity, we will often denote the full bispectrum as $B_{\bell \bfm}$ and the angle-averaged bispectrum as $b_{\bell}$.

\vskip 5pt
Applying the selection rule and the identity \eqref{eq:wig-parity}, the complex conjugate of the bispectrum \eqref{eq:red-bis-def} reads
\begin{align}
    B_{\bell \bfm}^* = (-1)^{m_{123}}\begin{pmatrix}
        \ell_1 & \ell_2 & \ell_3 \\
        -m_1 & -m_2 & -m_3
    \end{pmatrix} b_{\bell}  = (-1)^{\ell_{123}}\begin{pmatrix}
        \ell_1 & \ell_2 & \ell_3 \\
        m_1 & m_2 & m_3
    \end{pmatrix} b_{\bell}\;,
\end{align}
in which by comparison yields $b_{\bell}^* = (-1)^{\ell_{123}}b_{\bell}$. In other words, in contrast to the plane-parallel limit where the parity-even and odd parts are generally complex \cite{Jeong:2019igb,Paul:2024uim}, the full-sky bispectrum 
\begin{align}
    b_{\bell} \in \begin{cases}
        \R, & \text{for }\ell_{123} = \text{ even}, \\
        \ci\,\R, & \text{for }\ell_{123} = \text{ odd}.
    \end{cases}
\end{align}
Note that even in a parity-symmetric universe, bispectra can be either real or imaginary \cite{Meerburg:2016ecv,Jeong:2019igb,Coulton:2021yjq,Paul:2024uim}, decided by the intrinsic parity of the observables. The presence of parity-violating physics however, induces correlations in the other reality sector, which are otherwise vanishing in a parity-symmetric universe. We emphasise that the tactic for searching for parity violation is to devise estimators that \textit{isolate} the primordial parity-violating contributions. In the case of ${\ddm}$, we will see in section \ref{sec:harmonic-analysis} that the inclusion of an underlying parity-violating trispectrum leads to signals in the imaginary parts of $b_{\bell}$ and thus can be cleanly extracted through the statistic
\begin{align}\label{eq:b-po}
    b^{\mathsmaller{\rm PO}}_{\bell} = \frac{1}{2}\lr[{b_{\bell} - (-1)^{\ell_{123}}b_{\bell}}]\;,
\end{align}
or similarly with the full bispectrum $B_{\bell \bfm}$. In other words, we would like to show that the parity-violating signal in \ddm does not vanish in the statistic \eqref{eq:b-po}, whilst parity-symmetric signals do.

\subsubsection{Harmonic analysis}\label{sec:harmonic-analysis}

As mentioned in section \ref{sec:vel-fields}, the $\avg{\de\de v_\parallel}$ term does not receive contributions from the primordial parity-odd trispectrum, so we will concern ourselves only with the $\avg{\de\de[\de v_\parallel]}$ term. The projection of the density fields are simply
\begin{align}
    \de_{\ell m} &= \int \dd^2\nhat~Y^*_{\ell m}(\nhat) \int_0^\infty \dd\chi~W_\de(\chi)\de(\nhat\chi)\nn \\
    & = 4\pi \ci^{-\ell}\int \dd\chi~W_\de(\chi)\lr[{\int_\bk j_\ell(k\chi) Y^*_{\ell m}(\khat)\delta(\bk)}]\;, \label{eq:de-lm}
\end{align}
where we have denoted $W_\cO(\chi)$ as the radial window for the observable $\cO$. Fourier transforming the field $\de$, and using the Rayleigh expansion $\exp({-\ci \bk\vdot\nhat\chi}) = 4\pi \sum_{\ell m} \ci^{-\ell} j_\ell(k\chi)Y^*_{\ell m}(\khat)Y_{\ell m}(\nhat)$ with the orthogonality of spherical harmonics leads to the second line.  For the composite field $\De(\bx)\equiv\de(\bx) v_\parallel(\bx)$, this involves a convolution in Fourier space, i.e.
\begin{align}\label{eq:dv||-lm}
    \De_{\ell m} = \int \dd^2\nhat~Y^*_{\ell m}(\nhat) \int_0^\infty \dd\chi~W_\De(\chi)\int_{\bk,\bq}\ce^{-\ci\bk\vdot\nhat\chi}\lr[{\frac{\ci\nhat\vdot\qhat}{q}}]\de(\bk-\bq)\de(\bq)\;,
\end{align}
where the quantity in square brackets relates the velocity to the density fields through the continuity equation \eqref{eq:cont-eq}. Recognising that the $\nhat\vdot\qhat$ factor can also be expanded in terms of spherical harmonics 
\begin{align}
    \nhat\vdot\qhat = \frac{4\pi}{3}\sum_{M} Y_{1M}(\nhat)Y^*_{1M}(\qhat)\;,
\end{align}
we can simplify \eqref{eq:dv||-lm} to get
\begin{align}\label{eq:De-lm}
    \De_{\ell m} = \frac{(4\pi)^2}{3}(-1)^m\int_0^\infty \dd\chi~W_\De(\chi) \lr[{\sum_{L,M,M'} \cG_{1L\ell}^{M'M-m} \int_{\bk,\bq}j_L(k\chi) Y^*_{LM}(\khat)Y^*_{1M'}(\qhat)\de(\bk-\bq)\de(\bq)}]\;
\end{align}
by integrating over the LoS to yield the Gaunt symbol $\cG_{\ell_1\ell_2\ell_3}^{m_1m_2m_3}$ \eqref{eq:gaunt}. 

\vskip 5pt
As the heart of the following derivation relies only on the angular integrals in Fourier space within the square bracketed terms in \eqref{eq:de-lm} and \eqref{eq:De-lm}, we will suppress the radial integrals in the following discussion. The bispectrum $\avg{\de\de\De}$ in harmonic space contains
\begin{align}\label{eq:de-de-De}
    B_{\bell\bfm}^{\de\de\De} \supset
    \sum_{L,M,M'} \cG_{1L\ell_3}^{M'M-m_3}\int_{\{\khat_a\},\qhat} & Y^*_{\ell_1m_1}(\khat_1)  Y^*_{\ell_2m_2}(\khat_2) Y^*_{LM}(\khat_3)Y^*_{1M'}(\qhat)\avg{\de({\bk_1})\de({\bk_2})\de({\bq})\de({\bk_3-\bq})}\;.
\end{align}
The route to show that the angle-average of $B^{\de\de\De}_{\bell\bfm}$ automatically projects onto $\ell_{123}={\rm odd}$ for parity-odd primordial trispectra follows exactly what has been done for angular trispectra (see section 4 of \cite{Greco:2025xtt} for a worked example). Instead of deriving the angular trispectra for arbitrary parity-violating shapes, we will simply show that contributions from parity-symmetric and parity-violating trispectra separate into different reality sectors of $B^{\de\de\De}_{\bell\bfm}$. Taking the product of $\eqref{eq:de-de-De}$ with $(-1)^{\ell_{123}}$, we have
\begin{align}\label{eq:de-de-De-flipped}
    (-1)^{\ell_{123}}B_{\bell\bfm}^{\de\de\De} \supset
    \sum_{L,M,M'} \cG_{1L\ell_3}^{M'M-m_3}\int_{\{\khat_a\},\qhat} & Y^*_{\ell_1m_1}(\khat_1)  Y^*_{\ell_2m_2}(\khat_2) Y^*_{LM}(\khat_3)Y^*_{1M'}(\qhat)\nn\\
    &\times (-1)^{\ell_3+L+1} \avg{\de({-\bk_1})\de({-\bk_2})\de({-\bq})\de({-\bk_3+\bq})}\;,
\end{align}
where we have used $(-1)^\ell Y^*_{\ell m}(\khat) = Y^*_{\ell m}(-\khat)$ on all the spherical harmonics, and a subsequent change of variables on all the angular integrals pushes the minus signs into the trispectrum contribution. With the Gaunt symbol enforcing $\ell_3+L+1$ to be even, \eqref{eq:de-de-De-flipped} is exactly the same as \eqref{eq:de-de-De} with the trispectrum momenta flipped. Since parity-symmetric contributions to the density trispectrum obey
\begin{align}\label{eq:symm-trisp}
    \avg{\de({-\bk_1})\de({-\bk_2})\de({-\bq})\de({-\bk_3+\bq})} = \avg{\de({\bk_1})\de({\bk_2})\de({\bq})\de({\bk_3-\bq})}\;,
\end{align}
comparing \eqref{eq:de-de-De-flipped} with \eqref{eq:de-de-De}, it can be seen that the statistic
\begin{align}\label{eq:diff-stat}
    \frac{1}{2}\lr[{B^{\de\de\De}_{\bell\bfm}-(-1)^{\ell_{123}}B^{\de\de\De}_{\bell\bfm}}]
\end{align}
has no parity-symmetric contributions. Conversely, parity-violating signals change sign for an inversion of all the momenta and hence this bispectrum combination is non-vanishing. This confirms our expectations in section \ref{sec:sphere-symm}, that the parity-violating information can be isolated with the statistic above.

\vskip 5pt
As a final remark, one can also prove that the parity-violating signal does not vanish in the statistic $b_{\bell}^\mathsmaller{\rm PO}$ through the form given by  \eqref{eq:ddp-pv}. We show this derivation in appendix \ref{sec:ang-avg}.

\subsection{Parity from the sphere to the plane}

In parts \ref{sec:plane-parallel} and \ref{sec:projected-obs} above, we have encountered two different appearances of parity-violating contributions in {\ddm}. One is the factor of $\nhat\vdot(\khat_1\times\khat_2)$ in the plane-parallel limit of the Fourier space bispectrum, the other is the requirement on the sum of multipoles $\ell_{123}$ to be odd in the full-sky harmonic space bispectrum. We conclude this section with a discussion connecting the two.

\begin{figure}
    \centering
    \includegraphics[width=0.45\linewidth]{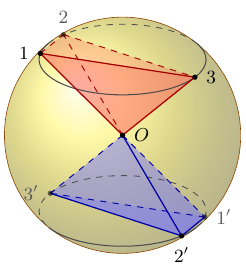}
    \caption{Sketch of three-point correlations on the sphere. The three points, labelled $1, 2, 3$, along with the origin $O$, form a non-planar tetrahedron, coloured red. Without loss of generality, we depict the points to be on a small circle on the northern hemisphere. Their parity-inverted counterparts, found by a point-inversion around $O$, are labelled $1',2',3'$, with the corresponding tetrahedron coloured blue. The two tetrahedra cannot be mapped onto each other by a proper rotation about $O$.}
    \label{fig:parity_sphere}
\end{figure}

\vskip 5pt
We restrict our discussion to projected fields as the parity-sensitivity of \ddm is already apparent in this setting. It is clear that the plane-parallel- and full-sky-bispectra are just different integral transforms of the correlation function $\avg{\de(\nhat_1)\de(\nhat_2)p_\parallel(\nhat_3)}$, where we note that
$p_\parallel(\nhat_3) = \nhat_3\vdot\bp(\nhat_3)$ in this context. For the harmonic space bispectrum, it is related to the angular space correlation function  by 
\begin{align}\label{eq:ang-corrfn}
    \avg{\de(\nhat_1)\de(\nhat_2)p_\parallel(\nhat_3)} &= \sum_{\ell_1,\ell_2,\ell_3}\sum_{m_1,m_2,m_3}B_{\bell\bfm} Y_{\ell_1m_1}(\nhat_1)Y_{\ell_2m_2}(\nhat_2)Y_{\ell_3m_3}(\nhat_3) \nn \\
    &=\sum_{\ell_1,\ell_2,\ell_3} b_{\bell}\sum_{m_1,m_2,m_3} 
    \begin{pmatrix}
        \ell_1 & \ell_2 & \ell_3 \\
        m_1 & m_2 & m_3
    \end{pmatrix}
    Y_{\ell_1m_1}(\nhat_1)Y_{\ell_2m_2}(\nhat_2)Y_{\ell_3m_3}(\nhat_3) \;. 
\end{align}
To see how the different reality sectors, or equivalently $\ell_{123}={\rm odd/even}$ parts, of $B_{\bell\bfm}$ (or $b_{\bell}$) contribute to the correlation function, we use the equivalence between spherical harmonics and rank-$\ell$ symmetric trace-free Cartesian tensors,
\begin{align}\label{eq:stf-decomp}
    Y_{\ell m}(\nhat) = \sum_{a_1,\ldots,a_\ell}\cY^{\ell m}_{a_1 \ldots a_\ell} \nh^{a_1}\ldots \nh^{a_\ell}\;.
\end{align}
Expanding all the spherical harmonics in \eqref{eq:ang-corrfn} as such, statistical isotropy demands that all the $\nhat_1$, $\nhat_2$, and $\nhat_3$'s must be contracted with the isotropic tensors $\de_{ij}$ or $\eps_{ijk}$. This immediately requires that non-vanishing parts of $B_{\bell\bfm}$ (or $b_{\bell}$) with $\ell_{123}={\rm odd}$ are proportional to $\nhat_1\vdot(\nhat_2\times\nhat_3)$. (This factor is related to the signed volume of the tetrahedron formed by the three points on the sphere along with the observer at the origin, which we sketch in figure \ref{fig:parity_sphere}.) Therefore, the correlation function can be written as 
\begin{align}\label{eq:pe-po-corr-decomp}
    \avg{\de(\nhat_1)\de(\nhat_2)p_\parallel(\nhat_3)} = \xi_{\mathsmaller{\rm PE}}(n_{12},n_{23},n_{31}) + [\nhat_1\vdot(\nhat_2\times \nhat_3)]\xi_{\mathsmaller{\rm PO}}(n_{12},n_{23},n_{31})\;,
\end{align}
where $n_{ij} = \nhat_i\vdot\nhat_j$, and  $\xi_{\mathsmaller{\rm PE/PO}}$ are scalar functions of the $n_{ij}$'s. Extracting the $\ell_{123}={\rm odd}$ part of the bispectrum, i.e.\ \eqref{eq:diff-stat}, is thus equivalent to searching for correlations on the sphere for which
\begin{align}
    \avg{\de(\nhat_1)\de(\nhat_2)p_\parallel(\nhat_3)} - \avg{\de(-\nhat_1)\de(-\nhat_2)p_\parallel(-\nhat_3)}\;,
\end{align}
 is non-zero, hence isolating $\xi_{\mathsmaller{\rm PO}}$. 

\vskip 5pt
In addition to the procedure outlined in section \ref{sec:plane-parallel}, the plane-parallel bispectrum can also be obtained through the correlation function \eqref{eq:pe-po-corr-decomp} by allowing the three LoS's to contract to a small patch of the sky around a central direction $\nhat$, which defines a tangent plane $T_{\nhat}S^2$ on the sphere. The plane-parallel approximation amounts to writing the individual LoS in a small-angle expansion \cite{Bernardeau:2010ac}
\begin{align}
    \nhat_i = \cos\th_i \nhat + \sin\th_i \ehat_{\perp,i}  = \nhat + \bth_i - \frac{1}{2}\th_i^2\nhat + \cO(\th_i^3) \;,
\end{align}
where $\th_i$ is the angle between $\nhat_i$ and the fixed direction $\nhat$, $\ehat_{\perp,i}$ is a unit vector in the tangent plane $T_{\nhat}S^2$, and we written $\bth_i \equiv \th_i \ehat_{\perp,i}$ in the latter equality. At leading order in this approximation, the rotational invariants are
\begin{align}
    \nhat_1\vdot(\nhat_2\times\nhat_3) &= \nhat \vdot \lr[{(\bth_2-\bth_1) \times (\bth_3 - \bth_1 )}] + \cO(\th_{1,2,3}^3)\;, \label{eq:scp-lim}\\
    \nhat_i\vdot\nhat_j  &= 1 - \frac{1}{2}|\bth_i-\bth_j|^2 +\cO(\th_{i,j}^3)\;. \label{eq:cos-lim}
\end{align}
Writing the Fourier transform in the tangent plane as 
\begin{align}\label{eq:2d-ft}
    \cO(\bk_\perp) = \int \dd^2\bth\ \ce^{\ci \bk_\perp\vdot\bth} \cO(\bth)\;, 
\end{align}
and $\bth_i$ factors as derivatives acting on the exponentials, one finds the $\xi_\mathsmaller{\rm PO}$ part of the correlation function gives a bispectrum
\begin{align}\label{eq:tns2-bis}
    \avg{\de(\bk_{\perp,1})\de(\bk_{\perp,2})p_\parallel(\bk_{\perp,3})} = (2\pi)^2 \de^{D}_{(2)}(\bk_{\perp,123})(\nhat\vdot(\bk_{\perp,1}\times\bk_{\perp,2}))B(k_{\perp,1},k_{\perp,2},k_{\perp,3})\;,
\end{align}
where $\de^D_{(2)}$ is the two-dimensional Dirac delta function on the tangent plane. This arises as a corollary of the leading order behaviour of \eqref{eq:scp-lim} and \eqref{eq:cos-lim} depending only on relative displacements $\bth_i-\bth_j$ in the tangent plane. This is the well-known result that rotational invariance emerges as translational invariance in the tangent plane at leading order in the plane-parallel limit \cite{Raccanelli:2023zkj,Gao:2023rmo}. (Radial differences between the points will appear as off-diagonal corrections if included.) As anticipated, \eqref{eq:tns2-bis}  reproduces the $\nhat\vdot(\bk_1\times\bk_2)$ structure we saw in equation \eqref{eq:ddp-pv}, as the radial Fourier components do not contribute in the triple scalar product. This alternative derivation also clarifies that the Fourier modes in the plane-parallel limit are defined locally with respect to the tangent plane, and any stacking or averaging procedure across multiple patches requires a joint rotation of the local vector $\nhat$ and the Fourier modes $\bk_i$.

\vskip 5pt
Observables for parity violation often derive their sensitivity from a pseudo-scalar, e.g.\ $EB$ in CMB polarisation \cite{Komatsu:2022nvu} and galaxy shapes \cite{Kurita:2025hmp}, and $\bv\vdot\bJ$ in halo spins \cite{Coulton:2023oug}, and the \ddm bispectrum is not an exception. 
As we have seen in this analysis, the parity-sensitivity of the plane-parallel- and the full-sky-bispectrum both stem from their dependence on the pseudo-scalar $\nhat_1\vdot(\nhat_2\times\nhat_3)$. Thus, the two approaches give consistent descriptions of the \ddm statistic in the appropriate limit.

\section{Case study: The kinetic Sunyaev-Zel'dovich effect}\label{sec:forecast}

To study the performance of $\avg{\de\de p_\parallel}$ as a probe of parity violation, we perform a Fisher forecast for the detectability of $\avg{\z^4}_\mathsmaller{\rm PO}$ under optimal weighting of a trispectrum template. The observable we have chosen to adopt as the momentum field is the secondary anisotropy induced in the CMB temperature  by the kinetic Sunyaev-Zel'dovich (kSZ) effect \cite{Sunyaev:1972eq,Sunyaev:1980vz}. These are imprints of incoming CMB photons that have elastically scattered off free electrons in bulk motion as they propagate through the large-scale structure of the Universe. The amount of Doppler shift in the CMB temperature, generated through the kSZ, is proportional to the momentum field of the electrons, i.e.\ \cite{Shao:2010md,Smith:2018bpn,Giri:2020pkk}
\begin{align}
    \De T_\mathsmaller{\rm kSZ}(\nhat) = \int \dd\chi~ W_T(\chi) \,p_\parallel(\nhat\chi)\;.
\end{align}
The window function $W_T(\chi)$ encodes the biased distribution of electrons, and the scattering rate as a function of comoving distance
\begin{align}
    W_T(\chi) = b_{\rm e}\s_{\rm T}\bar{n}_{\rm e,0}
    \big({1+z(\chi)}\big)^2\;,
\end{align}
with $b_{\rm e}$ the linear electron bias, $\s_{\rm T}$ the Thomson scattering cross section, $\bar{n}_{\rm e,0}$ is the mean physical electron number density at redshift zero, and $z(\chi)$ is the redshift as a function of comoving distance. We assume that the electron distribution is unbiased and take $ b_{\rm e}=1$ in this analysis. 

\vskip 5pt
 A direct map of the kSZ signal is still out of reach, and is technically challenging to obtain due to its spectral similarity with the primary blackbody spectrum of the CMB \cite{Foreman:2022ves,Raghunathan:2023yfe}. However, the kSZ effect has been detected to high significance through its cross-correlation with galaxy surveys through \ddm and its variants with galaxy clustering acting as the density legs \cite{Chaves-Montero:2019isa,Tanimura:2020une,Contreras:2022zdz,Krywonos:2024mpb,Lague:2024czc,Hotinli:2025tul,Lai:2025qdw,McCarthy:2025brx,Chaussidon:2026vmn,Qu:2026zyh}. As such, we choose this combination of probes for the Fisher forecast, in hope to piggyback onto current kSZ pipelines in future analyses. 

\subsection{Estimator and Fisher information}\label{sec:estimator}
To explore how well we can constrain the amplitude of the template in \eqref{eq:will-temp}, assuming that parity-violating effects are weak, we can make use of the extensive literature on optimal PNG amplitude estimators \cite{Babich:2005en,Smith_2015}. For simplicity, we assume that the noise covariance is diagonal, then we have
\begin{align}
    \hat{p}_\mathsmaller{\rm NL} = \frac{1}{2F}\sum_{\ell_{123}={\rm odd}}\sum_{m_1,m_2,m_3} \frac{\ptl B_{\bell\bfm}^*/\ptl \pnl}{C^{gg}_{\ell_1}C^{gg}_{\ell_2}C^{TT}_{\ell_3}}g_{\ell_1m_1}g_{\ell_2m_2} T_{\ell_3m_3} \;,
\end{align}
where $\avg{g_{\ell m }g_{\ell'm'}} = \de_{\ell\ell'}\de_{mm'}C_\ell^{gg}$ and $\avg{T_{\ell m }T_{\ell'm'}} = \de_{\ell\ell'}\de_{mm'}C_\ell^{TT}$ are the covariance matrices for the observed galaxy and CMB temperature field respectively, and the galaxy-temperature cross spectrum has been assumed to vanish. Note that we only sum over configurations where the sum of multipoles is odd, corresponding to the statistic \eqref{eq:diff-stat}, to isolate the contributions from PVICs. The prefactor in front takes care of the exchange symmetry between the two galaxy legs and  the normalisation of the estimator is
\begin{align}\label{eq:fish}
    F = \frac{1}{2}\sum_{\ell_{123}={\rm odd}}\sum_{m_1,m_2,m_3} \frac{\lvert \ptl B_{\bell\bfm}/\ptl \pnl \rvert^2}{C^{gg}_{\ell_1}C^{gg}_{\ell_2}C^{TT}_{\ell_3}}\;.
\end{align}
The estimator $\hat{p}_\mathsmaller{\rm NL}$ is unbiased and optimal with variance ${\rm Var} (\hat{p}_\mathsmaller{\rm NL}) = 1/F$. 

\vskip 5pt
For this analysis, we have chosen to perform the example for the primordial template \eqref{eq:will-temp}. The lack of dependence of the primordial template \eqref{eq:will-temp} on the diagonal momentum ($\bk_1 + \bk_2, \bk_1+\bk_3$) allows for the Fisher information $F$ to be recast in a factorisable form, allowing for efficient evaluation.\footnote{Trispectra with diagonal dependence but remain separable as in \cite{Lee:2020ebj} should still admit a factorisable Fisher information, but additional integrals stemming from momentum conservation across diagonals might render the computation  expensive \cite{Smith:2015uia, Philcox:2025bvj} but feasible for window functions with sufficiently narrow support \cite{Lee:2020ebj}.} The procedure mostly follows the algorithm for fast polyspectra estimation in the CMB \cite{Komatsu:2003iq,Smith:2006ud,Smith:2015uia, Duivenvoorden:2019ses,Philcox:2025bvj}, extended to angular projections of (composite) observables in large-scale structure. We spell out the general structure of the bispectrum and the Fisher information in the following and leave detailed derivations and interpretations to appendix \ref{sec:fact-fish}.

\vskip 5pt
For clarity, we will use labels $\ell_1, m_1,\ell_2, m_2$ for the galaxy multipoles, and labels $L,M$ for the temperature multipoles. The labels $\ell_3,m_3,\ell_4, m_4$ are instead reserved for the internal density and velocity legs making up the kSZ multipoles.
Given the trispectrum template \eqref{eq:will-temp}, the projected bispectrum takes the form 
\begin{align}\label{eq:B_ggT}
    B_{\ell_1\ell_2 L}^{m_1m_2M} &= -\ci\frac{9}{25}\pnl A^3\sum_{\tt ijk}\sum_{\tt abcd}\int\frac{\dd x}{2}\int_{\xhat} \int_{\chi_1} W_g(\chi_1)D(\chi_1) \int_{\chi_2} W_g(\chi_2)D(\chi_2) \int_\chi W_T(\chi) D^2(\chi) f(\chi)\cH(\chi)\nn\\
     &\times \eps_{\tt ijk}\eps_{\mathtt{abcd}}{\color{Blue3}\Big[{{^\mathtt{i}}G^\mathtt{a}_{\ell_1}(x,\chi_1) (-1)^{m_1} {_\mathtt{i}Y_{\ell_1 -m_1}}(\xhat)}\Big]}{\color{Red}\Big[{ {^\mathtt{j}}G^\mathtt{b}_{\ell_2}(x,\chi_2) (-1)^{m_2} {_\mathtt{j}Y_{\ell_2 -m_2}}(\xhat)}\Big]} \int_{\nhat}Y^*_{LM}(\nhat)\nn\\
    &\times 
    {\color{orange} \lr[{ \sum_{\ell_3,m_3} {^\mathtt{k}}G_{\ell_3}^\mathtt{c}(x,\chi)(-1)^{m_3} {_\mathtt{k}Y_{\ell_3-m_3}}(\xhat) Y_{\ell_3 m_3}(\nhat) }]}
    {\color{Green2} \lr[{ \sum_{\ell_4,m_4} V^\mathtt{d}_{\ell_4}(x,\chi)(-1)^{m_4} {Y_{\ell_4-m_4}}(\xhat) Y_{\ell_4 m_4}(\nhat) }]}\;,
\end{align}
where for brevity we write $A\equiv 2\pi^2 A_s$. $D(\chi)$ here represents the linear growth factor. The two Levi-Civita symbols encode the scalar triple product in $K_\mathsmaller{\rm PV}$ \eqref{eq:Kpv}, in which the indices are sampled from $\{0,+1,-1\}$ and the  permutations of the scalar bispectrum \eqref{eq:will-temp}, in which indices are taken from  $\{-5,-4,-3,0\}$.  Each coloured bracket represents the projection of a momentum leg of the primordial trispectrum to harmonic space, where $_{\tt s}Y_{\ell m }$ are spin-weighted harmonics, and the kernels are defined as 
\begin{subequations}\label{eq:GV-kernels}
\begin{align}
    {^{\tt s}}G^{\tt n}_\ell(x,\chi) &= (4\pi)^2 \L_\ell^{\tt s}\int \frac{k^2\dd k }{(2\pi)^3} j_\ell(k\chi)\cM_\de(k) k^{\tt n}\ptl_x^{1-|{\tt s}|} j_\ell(kx)  \;, \label{eq:G-x-chi}\\
    V^{\tt n}_\ell(x,\chi) &= (4\pi)^2\int \frac{k^2\dd k }{(2\pi)^3} \ptl_\chi j_\ell(k\chi)\cM_v(k) k^{\tt n} j_\ell(kx)\;,
\end{align}
\end{subequations}
with $\L^{\tt s}_\ell = {\tt s}\sqrt{\ell(\ell+1)}$ for ${\tt s} = \pm 1$ and $\L^{\tt s}_\ell = 1$ for ${\tt s}=0$. The $k$-dependent functions are
\begin{align}\label{eq:transfer}
    \cM_\de(k) = -\frac{2}{5}\frac{k^2 T(k)}{\O_{m,0} H_0^2}\;,\qquad \cM_v(k) = -k^{-2}\cM_\de(k)\;,
\end{align}
with $T(k)$ the curvature to density transfer function, normalised to unity at large scales. The integrals \eqref{eq:GV-kernels} can be evaluated efficiently by performing an FFTLog decomposition \cite{Hamilton:1999uv} of $\cM_\de(k)$ and the subsequent integral over Fourier modes $k$ can then be done analytically  \cite{Assassi:2017lea,Schoneberg:2018fis,Lee:2020ebj,Chen:2021vba}. Note that we never apply the Limber approximation throughout the numerical evaluation. More details are given in appendix \ref{sec:fftlog}. 

\vskip 5pt
The resulting Fisher information can be written in the concise form
\begin{align}
    F &=\frac{A^6}{8}  \lr({\frac{9}{25}})^2\int{\dd x}\int{\dd x'}\int_\chi W_T(\chi) D^2(\chi)f(\chi) \cH(\chi)\int_{\chi'} W_T(\chi') D^2(\chi')f(\chi') \cH(\chi') \nn\\
      &\hspace{2em}\times \sum_{\tt ijk}\sum_{\tt i'j'k'}\sum_{\tt abcd}\sum_{\tt a'b'c'd'}  \eps_{\tt ijk} \eps_{\tt i'j'k'}\eps_{\tt abcd} \eps_{\tt a'b'c'd'} \sum_L \frac{2L+1}{(4\pi)^3} \frac{1}{C^{TT}_L}{\color{Teal} G^\mathtt{aa'bb'}_{{\tt ii'jj'},L}}{\color{magenta}X^\mathtt{cd}_{{\tt k},L}}{\color{magenta}X^\mathtt{c'd'}_{{\tt k'},L}}\;,
\end{align}
where ${\color{Teal} G^\mathtt{aa'bb'}_{{\tt ii'jj'},L}}$ and ${\color{magenta}X^\mathtt{cd}_{{\tt k},L}}$ are projections of the colour-coded functions in \eqref{eq:B_ggT}. In particular, 
\begin{subequations}\label{eq:fact-kers}
\begin{align}
    {\color{Teal} G^\mathtt{aa'bb'}_{{\tt ii'jj'},L}(x,x')} &= \frac{1}{2}\int_{-1}^1 \dd\cos\th\  d^L_{\tt i+j,i'+j'}(\th) {\color{Blue3}\lr[{ \sum_{\ell_1} \frac{ {^\mathtt{i}}G^\mathtt{a}_{\ell_1} (x){^\mathtt{i'}}G^\mathtt{a'}_{\ell_1}(x')}{C_{\ell_1}^{gg}}  (2\ell_1+1) d^{\ell_1}_{\mathtt{i'i}}(\theta) }]} \nn\\
    &\hspace{11em}\times{\color{Red}\lr[{ \sum_{\ell_2} \frac{ {^\mathtt{j}}G^\mathtt{b}_{\ell_1} (x){^\mathtt{j'}}G^\mathtt{b'}_{\ell_1}(x')}{C_{\ell_2}^{gg}}  (2\ell_2+1)d^{\ell_2}_{\mathtt{j'j}}(\theta) }] }\;,\\
    {\color{magenta}X^\mathtt{cd}_{{\tt k},L}(x)}&= \frac{1}{2}\int_{-1}^1 \dd\cos\th ~d^L_{\tt k0}(\th) \int \dd\chi~ W_T(\chi) {\color{orange}\lr[{\sum_{\ell_3} {^{\tt k}G^\mathtt{c}_{\ell_3}(x,\chi)}  (2\ell_3+1 ) d^{\ell_3}_{\tt 0k}(\th) }]  } \nn\\
    &\hspace{14.5em}\times{\color{Green2} \lr[{ \sum_{\ell_4}{V^\mathtt{d}_{\ell_4}(x,\chi)}  (2\ell_4+1 )d^{\ell_4}_{\tt 00}(\th)  }] }\;,
\end{align}
\end{subequations}
where $d^{\ell}_{m,m'}(\theta)$ are Wigner $d$-matrices, defined in appendix \ref{sec:wig-d}. This form of the Fisher information, that factorises into different components, can be efficiently computed with a complexity of $\cO(N_\chi N_x^2  \ell_{\rm max}^{3})$, where $\ell_{\rm max}$ is the maximum multipole truncation of the inner sums in \eqref{eq:fact-kers}, and $N_x$ and $N_\chi$ are the numbers of quadrature points needed for the numerical integration over $x$ and $\chi$ respectively. This is a significant improvement over the naive computational cost of $\cO(\ell_{\rm max}^9)$ found by plugging in \eqref{eq:B_ggT} into \eqref{eq:fish}.\footnote{For those familiar with the optimal estimation of Fisher information for primordial trispectra in the CMB \cite{Smith:2015uia}, which boasts an improvement $\cO(\ell_{\rm max}^7)\to \cO(\ell_{\rm max}^2)$. The additional  $\ell_{\rm max}^2$ factor in the naive complexity derives its origin from the convolution of the density and velocity fields in the kSZ leg. The improvement in this work is to decrease this extra complexity to $\cO(\ell_{\rm max})$.} We use this factorisable form of the Fisher information for the numerical evaluation of all subsequent results, with vectorisation performed using \texttt{JAX} \cite{jax2018github}.

\subsection{Results}\label{sec:results}

In this section, we present the results of the Fisher forecast. For the baseline, we take the observed galaxies to have a constant linear bias $b_1=1.6$ and to be Gaussian distributed at a mean comoving distance of $\chi_0 = 2230$~Mpc (roughly $z_0\simeq 0.6$) with a width of $\s_\chi = 315$~Mpc (roughly $\s_z\simeq 0.1$). The angular shot noise is taken to be $N_\ell^{gg} = 4.5\times 10^{-7}$ per steradian. For the CMB, we have assumed a Gaussian beam and the noise covariance
\begin{align}
    N_\ell^{TT} = \De_T^2\,\ce^{\ell^2/\ell_b^2}\;
\end{align}
is taken to have detector noise $\De_T = 5~\mu\text{K-arcmin}$ and $\ell_b = \th_b^{-1}\sqrt{8\log 2} $, with $\th_b = 1.5$ arcmin, the full-width at half-maximum of the beam. The above values are chosen to be representative of a high number density bin in a SPHEREx-like galaxy survey \cite{SPHEREx:2014bgr} and CMB experiments with noise levels comparable to the Simons Observatory \cite{SimonsObservatory:2018koc}. The fiducial cosmology is assumed to be \LCDM with cosmological parameters taken from Planck \cite{Planck:2018vyg} calculated with ${\tt CAMB}$ \cite{Lewis:1999bs}.

\begin{figure}
    \centering
    \includegraphics[width=0.55\linewidth]{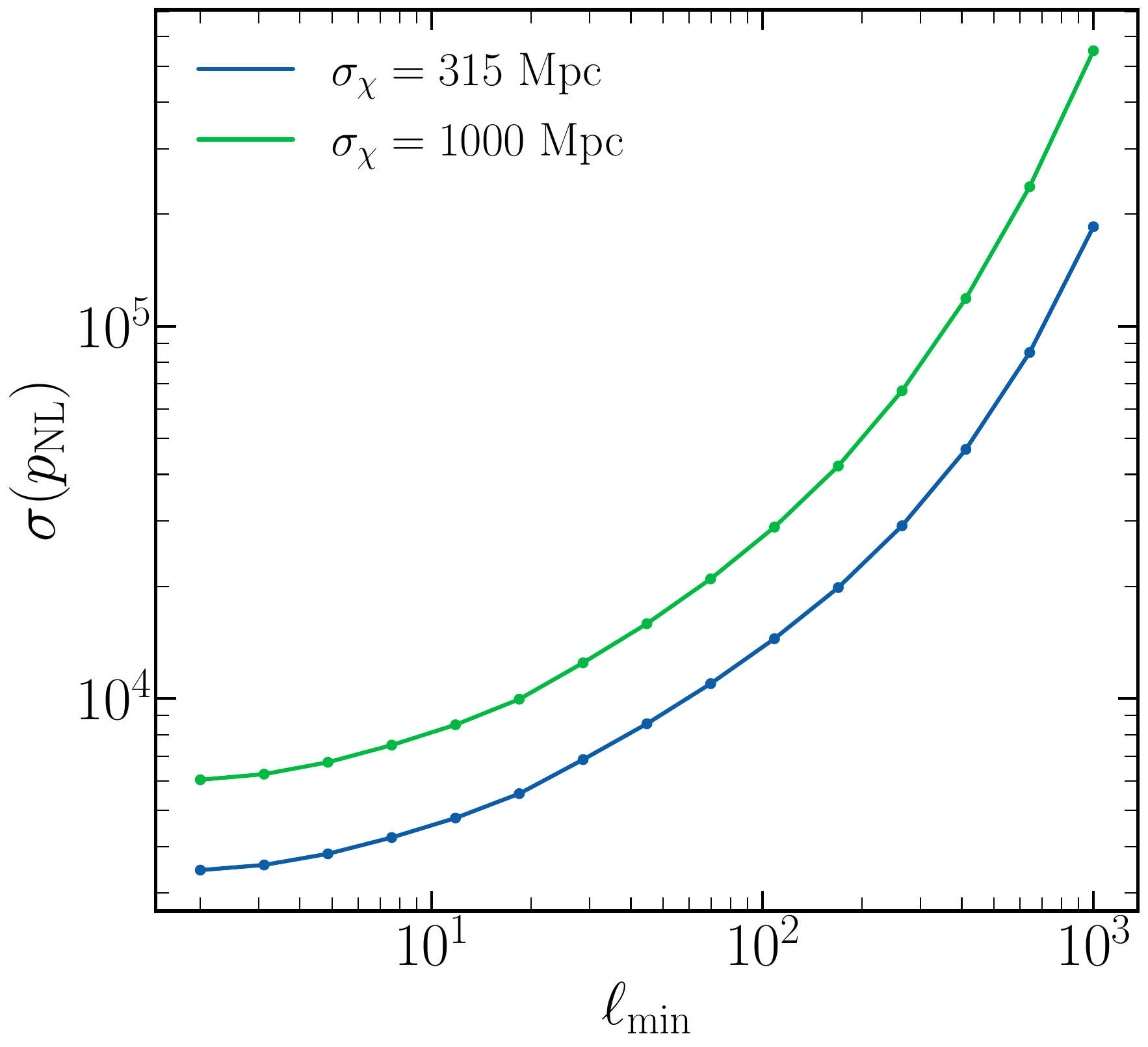}
    \caption{Forecast constraints on the the ultra-squeezed template amplitude as a function of $\ell_{\rm min}$ for Gaussian bin widths of $\s_\chi = 315$~Mpc (approximately $\s_z = 0.1$) and $\s_\chi = 1000$~Mpc (approximately $\s_z = 0.3$). Both configurations have a fiducial shot noise of $N^{gg}_\ell = 4.5\times 10^{-7}$/sr.}
    \label{fig:lmin}
\end{figure}

\vskip 5pt
In figure \ref{fig:lmin}, we first show the forecast constraints on the parity-violating amplitude as a function of the minimum multipole $\ell_{\rm min}$, which we chose to apply homogeneously for both the galaxy and temperature fields. The maximum multipole for both is fixed at $\ell_{\rm max}=5000$. Since the variance in the temperature is dominated by the primary CMB at large scales, the main effect of reducing $\ell_{\rm min}$ is to improve the measurement of the large scales in the galaxy survey alone. As the template constrained here peaks in the ultra-squeezed limit, our constraints improve as we minimise $\ell_{\rm min}$ and begin including the most squeezed configurations with the highest signals. Observational systematics such as galactic foregrounds can be significant at large scales but are not accounted for in this forecast. 

\vskip 5pt
To understand the effects of varying bin widths, we have also included the projected constraints for a redshift bin of Gaussian width $\s_\chi = 1000$~Mpc (roughly $\s_z = 0.3$) in figure \ref{fig:lmin}. The angular shot noise is kept to be the same at $N_\ell^{gg} = 4.5\times 10^{-7}$/sr to allow for a direct comparison. The overall $\ell_{\rm min}$-dependence is similar to the narrow bin case, but with about 2 -- 3 times poorer constraints. The reason for this degradation can be traced back to the use of velocity fields. As the radial bin broadens, it becomes significantly larger than the coherence length of the velocity field ($\De\chi\sim100$~Mpc \cite{Planck:2015ywj}). Then, the signal from the kSZ effect along the LoS \cite{McCarthy:2024nik} gets increasingly washed out. This can be avoided by including numerous narrow redshift bins and adopting a tomographic analysis. Fortunately, such procedures are already routine in modern kSZ analysis pipelines, and could be easily modified for parity violation searches. 

\begin{figure}
    \centering
    \begin{minipage}{0.495\textwidth}
            \includegraphics[width=0.99\linewidth]{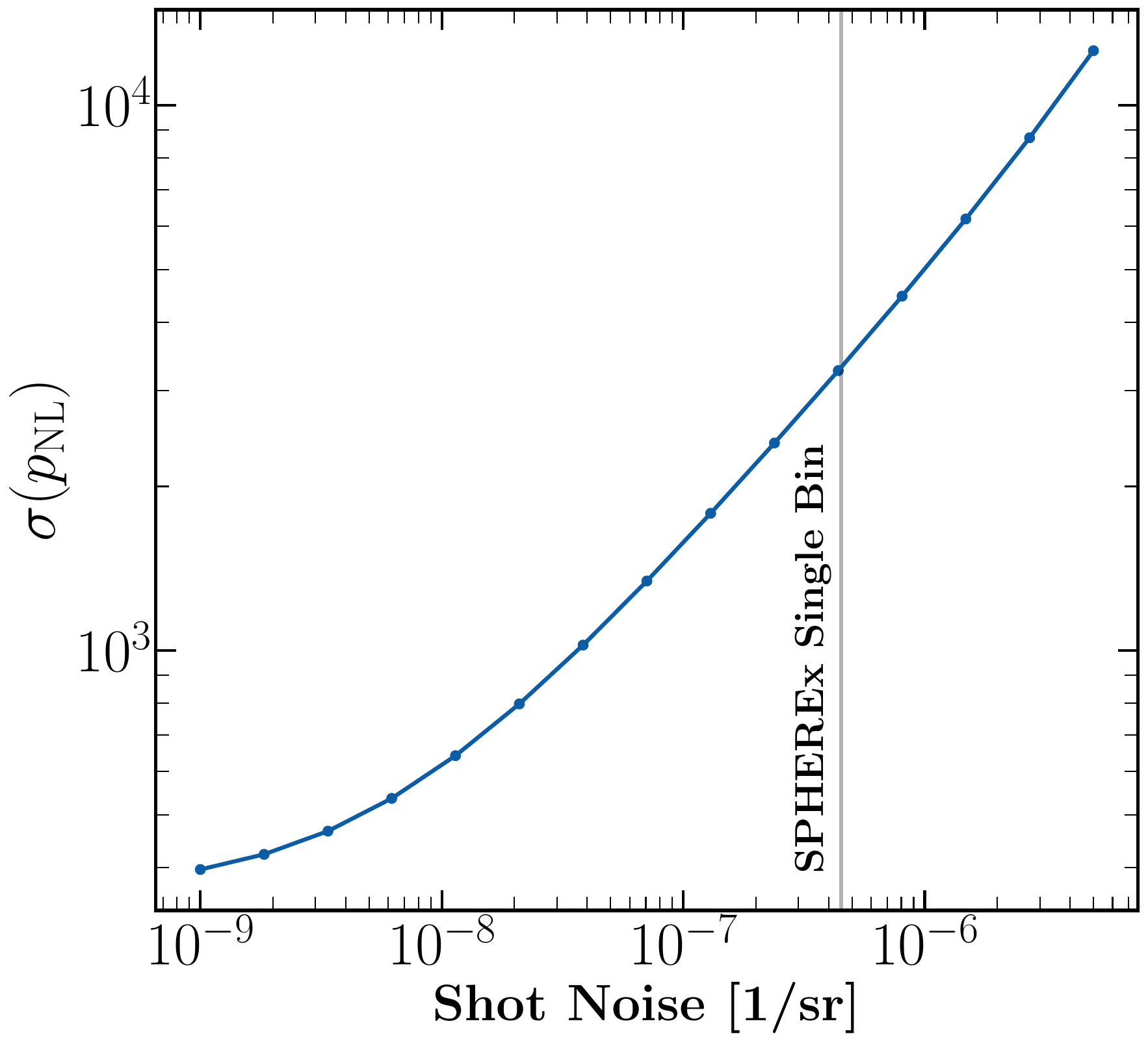}
    \end{minipage}
    \begin{minipage}{0.495\textwidth}
            \includegraphics[width=0.99\linewidth]{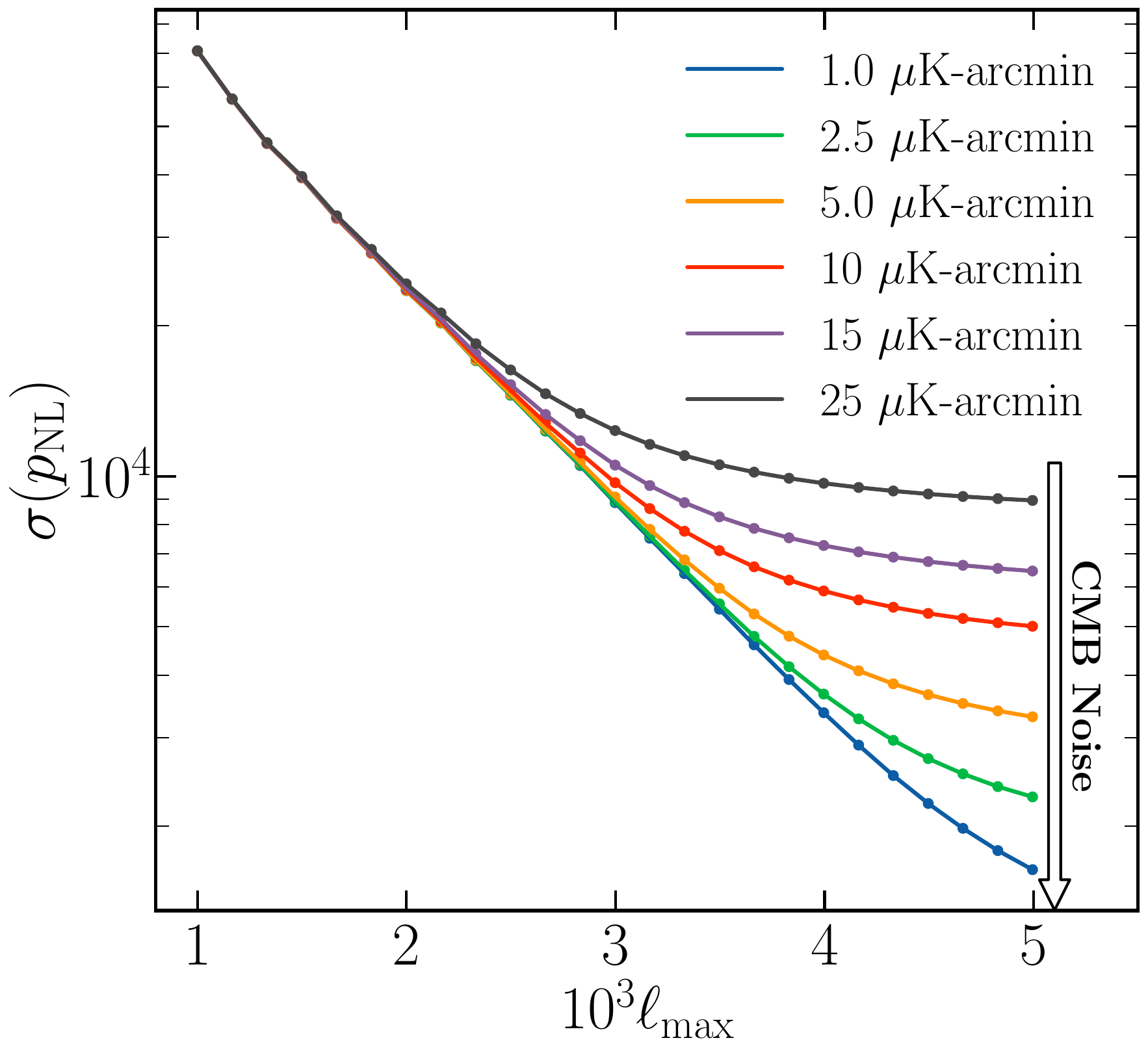}
    \end{minipage}
    \caption{\textit{Left: }Forecast constraints as a function of shot noise. The grey line indicates $N^{gg}_\ell = 4.5\times 10^{-7}$/sr, which we take as a typical noise level of a high number density SPHEREx bin. \textit{Right: }Forecast constraints as a function of the scale cut $\ell_{\rm max}$ for different levels of detector noise.}
    \label{fig:lmax}
\end{figure}

\vskip 5pt
On small scales, the main contributions to the variance are the shot noise in galaxy surveys, and the detector noise and foreground power in the CMB, and their impact on the forecast constraints are jointly shown in figure \ref{fig:lmax}. Note that due to the parity odd nature of our signal CMB foregrounds only act as a source of noise and do not bias this signal. For the shot noise plot, the maximum multipole is fixed at $\ell_{\rm max} = 5000$. We can see that the constraints get significantly better as a reduction in shot noise allows smaller modes to become resolved. This yields an increase in the number of well-measured independent modes, especially in the high-signal-ultra-squeezed-configurations important for the forecast. This continued improvement in constraining power as $\ell_{\rm max}$ is increased, as illustrated in the right panel of figure \ref{fig:lmax}, can be understood through the tensorial structure of the bispectrum in the plane-parallel limit, shown in \eqref{eq:ddp-pv} and \eqref{eq:tns2-bis}. In the Limber approximation where we only measure angular modes, the triple scalar product can be approximated as  $\nhat\vdot(\bk_{\perp,1}\times\bk_{\perp,2}) \simeq \bell_1\times\bell_2/\chi^2$ and does not vanish. This is in contrast to the behaviour of parity-symmetric contributions, because their signal is proportional to $\nhat\vdot\bk_i$, which is small in projection, i.e.\ $\nhat\vdot\bk_i\sim \nhat\vdot\bell_i \simeq 0$. This case demonstrates that sensitivity to PVICs can persevere in the high-$\ell$ limit, making  \ddm a promising statistic for future constraints on parity violation.

\section{Comparison with other approaches}\label{sec:discuss}
The characteristic structure of parity-violating signals, including their dependence on scalar triple products and the substantial cancellations that arise upon summing or integrating over configurations, complicates direct benchmarking of the sensitivity achieved by different observables and estimators relative to their parity-symmetric counterparts. This has garnered growing efforts \cite{Hou:2025cey,Kurita:2025hmp,Gao:2025yqd,Chen:2026wok} to forecast constraints on parity-violating amplitudes with a range of summary statistics, estimator constructions, and analysis frameworks. To contextualise the sensitivity obtained in our analysis, we compare our projected constraints from the galaxy-galaxy-kSZ bispectrum with those reported there. Note that different forecasts make different simplifying assumptions, and a direct comparison is not straightforward. We hereby provide some heuristic arguments on where our approach overlaps and differs from previous work, and hope this comparison provides a useful basis for identifying the relative strengths and limitations of the various approaches.

\vskip 5pt
To reiterate, there are two main motivations in incorporating the density-density-momentum bispectrum into the toolkit for parity-violating searches. One prime motivation is to exploit lower-order statistics to utilise the full potential of the data at hand. At this front, our work is spiritually similar to that of compressed statistics with composite fields in \cite{Jamieson:2024mau,Gao:2025yqd,Gao:2026zaz}. With a Gaussian smoothing scale of $5$~Mpc$/h$ (roughly $k_{\rm max} \simeq 0.2\ h/$Mpc), the authors of \cite{Gao:2025yqd} were able to constrain the parity-violating amplitude to $\s(\pnl) \simeq 8\times 10^4$ in a 1 $({\rm Gpc}/h)^3$ volume.\footnote{We translated the measurement of the maximum signal-to-noise (SNR) from \cite{Gao:2025yqd}  obtained in the \texttt{Quijote-ODD} simulations to a projected constraint by estimating $\s(\pnl) = |\pnl|/{\rm SNR}$, as this offers a more direct comparison to our Fisher constraints than the $z$-scores reported there.} In comparison, we achieve a similar constraint at $\ell_{\rm max}\simeq 10^3$ (roughly $k_{\rm max} \simeq 0.65\ h/$Mpc) in a volume of about  20 $({\rm Gpc}/h)^3$. 

\vskip 5pt
The discrepancy in the performance can be attributed to the retention of 3D information in the design of composite field estimators, where one can fully control their optimality. In our case, both the non-optimal compression of information in the momentum field as well as the use of 2D observables like the kSZ allows for less flexibility in estimator optimisation and hence degraded constraining power. As such, it would be interesting to perform the forecast for 3D observables of the momentum field, and understand if the information loss is indeed a result of the 3D-to-2D projection or a genuine shortcoming of the momentum field as a probe.  The merit of using momentum fields however lies at small scales, since information can still be extracted in that regime through the density-density-momentum bispectrum. This is especially true for kSZ anisotropies, whose effects relative to the primary CMB are most pronounced at small scales. Note however, that a realistic assessment of this claim requires a robust accounting of nonlinearities, which, unlike \cite{Gao:2025yqd}, we have omitted in our paper. For the kSZ effect, this requires non-Gaussian simulations of the full sky which are under development \cite{Bayer:2024egi,Backlight:inprep}. On the other hand, we have only performed a forecast with a single galaxy bin, which does not make use of the wide redshift coverage of the kSZ effect. With minimal changes however, our method is readily adaptable to current kSZ pipelines and can be extended to a tomographic analysis, using which constraints may be improved. To develop, validate, and explore systematic biases for such pipelines is a significant task. Thus, being able to co-opt existing pipelines for the purpose of parity violation searches should be significantly easier than developing one from scratch. Besides the differences between the approach presented here and that of \cite{Jamieson:2024mau,Gao:2025yqd,Gao:2026zaz}, there are also powerful synergies. For example, the observed distribution of tracers show distortions which depend on $\ptl^m_\parallel[ (1+\de)v_\parallel^m]$ \cite{Seljak:2011tx,Okumura:2013zva,Sugiyama:2015dsa}, which introduces additional parity-violating corrections in redshift space similar to those analysed in this paper. We leave exploration of this effect to future work.

\vskip 5pt
The second motivation to this project was to expand the current set of observables beyond the galaxy four-point correlation function and its derivatives \cite{Cahn:2021ltp,Hou:2022wfj,Philcox:2022hkh,Adari:2024vkf}. This resembles recent attempts to probe parity violation through weak lensing of the CMB and galaxies \cite{Greco:2025xtt,Chen:2026wok}. In the signal-dominated limit, their results show that the projected constraints on the parity-violating amplitude for the ultra-squeezed template are independent of the source redshift. At an $\ell_{\rm max}$ of $100$, the authors forecast a constraint of $\s(\pnl)\simeq 3\times 10^5$.\footnote{The papers \cite{Greco:2025xtt,Chen:2026wok} adopt a different normalisation for the parity-violating amplitude, which they call $g_-$, to the used here. The conversion between the two is given by $\pnl = (3/5)^6 g_-$.} Extrapolation of their results to higher multipoles is not straightforward, although a naive attempt would suggest that the performance is mildly better than ours in the noise-free limit. The slight advantage again owes its origins to the freedom in choosing optimal filters for all four legs of the trispectrum. For the case of CMB lensing, it would be interesting to understand how the constraints depend on the small-scale cutoff in the presence of instrumental noise, which was neglected in the analysis of \cite{Greco:2025xtt}, allowing a more direct comparison with our work here.

\vskip 5pt
In addition to the comparisons discussed above, there have been many other constraints, both established and forecast, of parity-violating models and templates \cite{Cabass:2022oap,Philcox:2023ypl,Philcox:2025bvj,Philcox:2025wts,Kurita:2025hmp,Chen:2026wok}. These include, primarily, parity-violating templates of the collapsed type. Although we do not carry out a detailed analysis of these templates here, the framework outlined in this work could be naturally applied to probe parity-violating trispectra of various shapes, which would test the reach of the density-density-momentum bispectra across a broader range of parity-violating scenarios.

\section{Summary and outlook}\label{sec:conclude}
In this work, we have proposed a novel class of observables, density-density-momentum bispectra, as complementary probes of parity violation. The major motivation behind this project is to search for complementary probes and statistics to the conventional galaxy four-point correlation function. As peculiar velocity measurements become more refined, with methods that are independent of the underlying density fields, correlations of the velocity field could provide alternative information to augment current parity violation searches. The innate nonlinearity tied to the sparse sampling of peculiar velocities allows the effective observable, the momentum field (i.e.\ the density-weighted velocity), to be sensitive to parity-violating effects already at the bispectrum level. The difference in the dimensionality of data vectors relative to a four-point-based approach reduces the computational challenges faced when using higher-order statistics.

\vskip 5pt
 We have shown that the non-vanishing transverse components of the momentum field are crucial for parity-violating signatures to be present in the density-density-momentum bispectrum, inducing correlations sensitive to the  orientation of the bispectrum triangle with a normal $\bk_1\times \bk_2$, a tell-tale sign of parity-violating physics. We find that the signal persists at the level of projected observables, such as CMB secondaries, and that it can be isolated from parity-symmetric contributions in \LCDM by selecting bispectrum configurations where the sum of multipoles is odd.

\vskip 5pt
To investigate the efficacy of \ddm in constraining parity violation, we performed a Fisher forecast for the parity-violating amplitude of the ultra-squeezed shape with galaxy clustering and kinetic Sunyaev-Zel'dovich anisotropies as the observables for the density and momentum fields respectively. With SPHEREx-like number densities and Simons Observatory-like noise levels, we report that the ultra-squeezed amplitude can be constrained to $\sigma(\pnl)\simeq\cO(10^{4})$ with a single galaxy bin, offering competitive constraints to other probes of parity violation. While we have opted for a template-driven forecast with specific tracers of the density and momentum field, our approach is generalisable to template-agnostic searches and across a wide range of cosmological observables, vastly expanding the scope of parity-violating searches.

\vskip 5pt
Our work has established a new window into parity violation, and suggests some theoretical and observational avenues for investigation:
\begin{itemize}
    \item \textbf{Redshift-Space Tracer Bispectra: }Although we have mainly focussed our attention to direct probes of the velocity field, we believe that similar signatures should be present in the redshift-space bispectrum of biased tracers. They should allow one to extract parity-violating information from galaxy surveys through the bispectrum alone.
    Moreover, $N$-body simulations with PVICs \cite{Coulton:2023oug} are publicly available, from which contributions from \ddm to the redshift-space bispectrum can be directly extracted.
    \item \textbf{Template Banks: }Model-agnostic estimators are suboptimal. It has been demonstrated that template-informed estimators can outperform fully model-agnostic searches by a factor of $2\sim3$ \cite{Gao:2025yqd}. Fourier-space-based estimators such as those presented in this work are well-suited for template-based searches. As such, exploring the theory space for PVICs and generating templates that generically capture their parity-violating behaviour would be very useful. As different observables are suited to capturing different shapes, this would also enable efficient benchmarking of the shape-sensitivity of different approaches.
    \item \textbf{Nonlinearities and Renormalisation: }Backreaction from nonlinearities in short modes will modify the shape of the parity-violating signals at quasi-linear scales and induce mode coupling, rendering current methods employed in parity-violating searches less effective. The effects of nonlinearities have been neglected in the estimation of both our signal and covariance. One way to include such effects is through dedicated simulations for parity violation. One could also proceed with analytical methods, along the lines of what was done for the matter trispectrum in the work of \cite{Azyzy:2025zif}, or that of \cite{Raikman:2026gcx} for the \ddm statistic with Gaussian initial conditions. Additionally, a similar treatment for biased tracers would also inform us about the new operators to include in the bias expansion in the presence of PVICs (only one of which has been currently identified \cite{Kurita:2025hmp}).
    \item \textbf{Skew-Spectra: }Current measurements of the galaxy-galaxy-kSZ cross-correlation are performed using a weighted skew-spectrum (also called a quadratic estimator in the CMB literature) instead of a maximum likelihood estimator for the bispectrum amplitude, although the two are closely related \cite{Contreras:2022zdz,McCarthy:2024nik,Lai:2025qdw}. The current skew-spectrum / quadratic estimator construction involves a weighted product in pixel space, which singles out only parity-symmetric contributions. To fully adapt the \ddm bispectrum for current pipelines, it would need to be recast in a skew-spectrum form, allowing constraints to be obtained from current data.
    \item \textbf{Momentum Field Power Spectrum: }It is also interesting to consider the momentum field power spectrum for probing parity violation if one is able to measure all three components of the momentum field. In general, the momentum field power spectrum $\avg{\bp(\bk)\otimes\bp(\bk')}$ can be written as \cite{Vlah:2019byq}
    \begin{align}
         (2\pi)^3\de^D(\bk+\bk')\lr[{ P_0(\bk)(\khat\otimes\khat) + P_+(\bk) \lr({\mathds{1}-\khat\otimes\khat}) + \ci P_\times(\bk) \lr({\boldsymbol{\eps}\vdot\khat})}]\;,
    \end{align}
    where $(\boldsymbol{\eps}\vdot\khat)_{ij} = \eps_{ijm}\hat{k}_m$. The parity-violating trispectrum can contribute to the parity-odd component $P_\times(\bk)$, which is only constrained to be zero by symmetry if projected along the line-of-sight. With the recent detection of the moving-lens effect \cite{Hotinli:2026wly}, the utility of this statistic for parity violation searches should be reevaluated. It would be interesting to understand the extent to which reconstructed velocity components from both the kSZ effect and the moving-lens effect can be combined to provide constraints on PVICs.
\end{itemize}
Parity violation is an intriguing possibility of inflationary microphysics, and as we have discussed, much work is necessary to take parity-violating searches to an equal footing with conventional searches for primordial non-Gaussianity. Towards this goal, our work has presented a new viewpoint into parity violation probes through the use of momentum fields as a composite observable. Ongoing peculiar velocity surveys and CMB experiments will deliver high quality maps that will improve our measurements of large-scale velocity fields in our Universe, allowing for new tests of parity violation in our Universe.

\acknowledgments 
The authors thank Azadeh Moradinezhad Dizgah, Pritha Paul, Xi Tong and Zvonimir Vlah for many orienting discussions, and Eiichiro Komatsu for comments on the draft. YSY is especially thankful to Matthew Johnson for his introduction to the kSZ effect, which was the inspiration for this work. YSY is grateful for the hospitality of the NCTS-Hsinchu Hub and National Tsing Hua University for their hospitality while parts of this work were being performed. YSY is supported by the Isaac Newton Studentship of the Cambridge Trust. WRC is supported by the Science and Technology Facilities Council (STFC) through an Ernest Rutherford Fellowship (UKRI2424: “Astrophysics and Cosmology from CMB Scattering”), through the Royal Society grant ICA\textbackslash R2\textbackslash 252140 and is grateful for past support from a KICC Kavli Fellowship. JH is grateful to Will Handley for his support. JH acknowledges funding by the Deutsche Forschungsgemeinschaft (DFG) through the Emmy Noether Programme -- Project number 554476934.  This research was supported in part by grant NSF PHY-2309135 to the Kavli Institute for Theoretical Physics (KITP).

This work used the DiRAC Memory Intensive service (Cosma8) at Durham University, managed by the Institute for Computational Cosmology on behalf of the STFC DiRAC HPC Facility (\url{www.dirac.ac.uk}). The DiRAC service at Durham was funded by BEIS, UKRI and STFC capital funding, Durham University and STFC operations grants. DiRAC is part of the UKRI Digital Research Infrastructure. This work made use of ChatGPT (GPT-5.5) by OpenAI for literature review and translation of pre-existing \texttt{Python} code into \texttt{JAX}. Underlying computational methodologies were developed independently of the AI-assisted translation. The authors take full responsibility for the content of this manuscript.

\appendix

\section{Wigner symbols and matrices}\label{sec:wig}
In this appendix, we summarise properties of Wigner symbols and matrices that are relevant throughout this work. 

\subsection{Wigner symbols}
Wigner symbols guide the decomposition of the tensor product space $\cV_{\bell_1}\otimes \cV_{\bell_2}\sim \bigoplus_{\bell} N_{\bell} \cV_{\bell}$ along unitary irreducible representations (UIRs) of Lie groups. Focussing on SO(3), the Wigner 3j symbols are the coefficients
\begin{align}\label{eq:wig3j}
    \begin{pmatrix}
        \ell_1 & \ell_2 & \ell \\
        m_1 & m_2 & m
    \end{pmatrix} \equiv \frac{(-1)^{\ell_1-\ell_2-m}}{\sqrt{2\ell+1}}\langle \ell_1 , m_1;\ell_2,  m_2\vert \ell,-m\rangle\;.
\end{align}
With the definition \eqref{eq:wig3j}, the orthogonality in the tensor product states is inherited by the 3j symbols in two relations 
\begin{subequations}
\begin{align}
    \sum_{\ell,m}(2\ell+1) \begin{pmatrix}
        \ell_1 & \ell_2 & \ell \\
        m_1 & m_2 & m
    \end{pmatrix} \begin{pmatrix}
        \ell_1 & \ell_2 & \ell \\
        m_1' & m_2' & m
    \end{pmatrix} &= \de_{m_1m_1'}\de_{m_2m_2'}\;,\\
    \sum_{m_1,m_2}(2\ell+1) \begin{pmatrix}
        \ell_1 & \ell_2 & \ell \\
        m_1 & m_2 & m
    \end{pmatrix} \begin{pmatrix}
        \ell_1 & \ell_2 & \ell' \\
        m_1 & m_2 & m'
    \end{pmatrix} &= \de_{\ell\ell'}\de_{mm'}\;. \label{eq:ortho-m}
\end{align}
\end{subequations}
The 3j symbols \eqref{eq:wig3j} vanish unless they satisfy selection rules:
\begin{align}\label{eq:select-rules}
    &|m_i|\leq \ell_i\;,\;~ |m|\leq \ell\;,\;~ m_1 + m_2 + m = 0 \;,\;~|\ell_1-\ell_2| \leq \ell \leq \ell_1+\ell_2\;.
\end{align}
Additionally, the Wigner 3j symbols possess many symmetries. Under column permutations, the 3j symbols transform as
\begin{align}
    \begin{pmatrix}
        \ell_{\s(1)} & \ell_{\s(2)} & \ell_{\s(3)} \\
        m_{\s(1)} & m_{\s(2)} & m_{\s(3)}
    \end{pmatrix} = {\rm sgn}(\s)^{\ell_1+\ell_2+\ell_3}
    \begin{pmatrix}
        \ell_1 & \ell_2 & \ell_3 \\
        m_1 & m_2 & m_3
    \end{pmatrix}\;,
\end{align}
where $\s\in S_3$, the symmetric group on three elements. Under simultaneous flipping of the $m$ labels, the symbols are related by a phase
\begin{align}\label{eq:wig-parity}
    \begin{pmatrix}
        \ell_1 & \ell_2 & \ell \\
        -m_1 & -m_2 & -m
    \end{pmatrix} = 
    (-1)^{\ell_1+\ell_2+\ell}\begin{pmatrix}
        \ell_1 & \ell_2 & \ell \\
        m_1 & m_2 & m
    \end{pmatrix}\;.
\end{align}
The Wigner 3j symbols are further related to spin-weighted spherical harmonics by the integral
\begin{align}\label{eq:gaunt}
    {}_{s_1s_2s_3}\cG_{\ell_1\ell_2\ell_3}^{m_1m_2m_3} &\equiv \int \dd^2\nhat ~{}_{s_1}Y_{\ell_1 m_1}(\nhat){}_{s_2}Y_{\ell_2 m_2}(\nhat){}_{s_3}Y_{\ell_3 m_3}(\nhat) \\
    &= \sqrt{\frac{(2\ell_1+1)(2\ell_2+1)(2\ell_3+1)}{4\pi}}
    \begin{pmatrix}
        \ell_1 & \ell_2 & \ell_3 \\
        -s_1 & -s_2 & -s_3
    \end{pmatrix} 
    \begin{pmatrix}
        \ell_1 & \ell_2 & \ell_3 \\
        m_1 & m_2 & m_3
    \end{pmatrix}\;.
\end{align}
The special case where $s_1=s_2=s_3 = 0$ gives the definition for the Gaunt symbol $\cG_{\ell_1\ell_2\ell_3}^{m_1m_2m_3}\equiv {}_{000}\cG_{\ell_1\ell_2\ell_3}^{m_1m_2m_3}$, which enforces a parity condition  $\ell_1+\ell_2+\ell_3  = {\rm even}$. This also gives a relation for the product of two spin-weighted spherical harmonics 
\begin{align}\label{eq:swsh-prod}
    {}_{s_1}Y_{\ell_1 m_1}{}_{s_2}Y_{\ell_2 m_2} = \sum_{\ell_3,m_3}\sqrt{\frac{(2\ell_1+1)(2\ell_2+1)(2\ell_3+1)}{4\pi}}
    \begin{pmatrix}
        \ell_1 & \ell_2 & \ell_3 \\
        -s_1 & -s_2 & -s_3
    \end{pmatrix} 
    \begin{pmatrix}
        \ell_1 & \ell_2 & \ell_3 \\
        m_1 & m_2 & m_3
    \end{pmatrix} {}_{s_3}Y^*_{\ell_3m_3}\;.
\end{align}
\subsection{Wigner $d$-matrix}\label{sec:wig-d}
The (reduced) Wigner $d$-matrices encode the matrix elements of a rotation $\b$ about the $y$-axis in the angular-momentum basis, i.e.
\begin{align}
    d^\ell_{m,m'}(\b) \equiv \langle \ell,m\vert\exp(-\ci\b L_y) \vert\ell, m'\rangle\;, 
\end{align}
where $L_y$ is the generator of $SO(3)$ along the $y$-axis. For integer values of $m$ and $m'$, the Wigner $d$-matrices satisfy the orthogonality relation
\begin{align}
    \int_{-1}^1\dd(\cos\b)~ d^{\ell}_{m,m'}(\b)d^{\ell'}_{m,m'}(\b) = \frac{2\de_{\ell\ell'}}{2\ell+1}\;.
\end{align}
The Wigner $d$-matrices also satisfy the symmetries
\begin{align}\label{eq:wig-d-symms}
    d^\ell_{m,m'}(\b) = d^\ell_{-m',-m}(\b) = (-1)^{m'-m} d^\ell_{m',m}(\b)\;.
\end{align}
Wigner $d$-matrices are tied to spin-weighted spherical harmonics by the relation
\begin{align}
    _{s}Y_{\ell m}(\th,\f) = (-1)^s \sqrt{\frac{2\ell+1}{4\pi}} \ce^{-\ci m\f}d^{\ell}_{m,-s}(\th)
\end{align}
which allows \eqref{eq:gaunt} to also be expressed as 
\begin{align}\label{eq:triple-d}
    \int_{-1}^1 \dd(\cos\th)~d^{\ell_1}_{m_1,s_1}(\th)d^{\ell_2}_{m_2,s_2}(\th)d^{\ell_3}_{m_3,s_3}(\th) = 2
    \begin{pmatrix}
        \ell_1 & \ell_2 & \ell_3 \\
        m_1 & m_2 & m_3
    \end{pmatrix}
    \begin{pmatrix}
        \ell_1 & \ell_2 & \ell_3 \\
        s_1 & s_2 & s_3
    \end{pmatrix} \;.
\end{align}

\section{No parity violation in longitudinal modes}\label{sec:deriv}
In this appendix, we prove that PVICs do not contribute to the $\avg{\de\de v_\parallel}$ term in \eqref{eq:ddp} to all loop orders in SPT. In SPT, the velocity field can be fully expressed in terms of a longitudinal mode. This is usually written in terms of the velocity divergence, $\bv(\bk) \equiv \ci\bk\theta(\bk)/k^2$. Expanded perturbatively, the $n$-th order contribution to the velocity field takes the form
\begin{align}
    \bv^{(n)}(\bk) = \ci f\cH\frac{\khat}{k}\int_{\{\bq_a\}}(2\pi)^3\de^D(\bk-\bq_{1\ldots n})G_n(\bq_1,\ldots,\bq_n)\prod_{a=1}^n\de^{(1)}(\bq_a)\;,
\end{align}
with $G_n$ the standard SPT kernels for velocity fields (See \cite{Bernardeau:2001qr} for a review on SPT). Note that the $\khat$ factor ensures that the velocity field is irrotational at all orders ($\nabla\times \bv = 0$) in SPT. We now consider the arbitrary loop contribution to the bispectrum
\begin{align}\label{eq:loop-ddv}
    \avg{\de^{(n_1)}({\bk_1})\de^{(n_2)}({\bk_2}) v_\parallel^{(n_3)}({\bk_3})} = (2\pi)^3\de^D(\bk_{123}) B^{n_1n_2n_3}_{\bk_1,\bk_2,\bk_3}\;.
\end{align}
At this order in perturbation theory, the loop contribution is
\begin{align}
    B^{n_1n_2n_3}_{\bk_1,\bk_2,\bk_3} &= (2\pi)^9\ci f\cH\frac{\mu_3}{k_3}\int_{\{\bq_a\}} F_{n_1}(\{\bq_\a\})F_{n_2}(\{\bq_\b\})G_{n_3}(\{\bq_\g\})\avg{\prod_{a=1}^{n_{123}} \de^{(1)}(\bq_a)}'\nn \\
    &\hspace{7em} \times \de^D\lr({\bk_1 - \sum_{a\in\a}\bq_a})\de^D\lr({\bk_2 - \sum_{a\in\b}\bq_a})\de^D\lr({\bk_3 - \sum_{a\in\g}\bq_a})\;,
\end{align}
with $F_n$ the standard SPT kernel for the $n$-th order contribution to the density field $\de^{(n)}$ \cite{Bernardeau:2001qr}. For brevity, we compile sets of  loop momenta as $\{\bq_\a\} = \{\bq_1,\ldots,\bq_{n_1}\}$, $\{\bq_\b\} = \{\bq_{n_1+1},\ldots,\bq_{n_{12}}\}$, and $\{\bq_\g\} = \{\bq_{n_{12}+1},\ldots,\bq_{n_{123}}\}$. We also denote $\mu_3 = \nhat\vdot \khat_3$ and $n_{abc} = n_a + n_b + n_c$. For the parity-violating contribution, correlators of the linear density field gain a minus sign when all momenta are flipped, i.e.
\begin{align}
    B^{n_1n_2n_3}_{\bk_1,\bk_2,\bk_3} &= -(2\pi)^9\ci f\cH\frac{\mu_3}{k_3}\int_{\{\bq_a\}} F_{n_1}(\{\bq_\a\})F_{n_2}(\{\bq_\b\})G_{n_3}(\{\bq_\g\})\avg{\prod_{a=1}^{n_{123}} \de^{(1)}(-\bq_a)}'_{\rm PV}\nn \\
    &\hspace{7em} \times \de^D\lr({\bk_1 - \sum_{a\in\a}\bq_a})\de^D\lr({\bk_2 - \sum_{a\in\b}\bq_a})\de^D\lr({\bk_3 - \sum_{a\in\g}\bq_a})\;.
\end{align}
The momentum-conserving delta function in \eqref{eq:loop-ddv} defines a plane with unit normal $\ehat$. Under the assumption of statistical isotropy, the correlation function inside the integrand is invariant under rotations of the internal momenta, i.e.\ $\avg{\prod _a\de(-\bq_a)} = \avg{\prod_a\de(-{\sf R}\bq_a)}$ for any rotation matrix ${\sf R}\in SO(3)$. Choosing ${\sf R}$ to represent a $180^\circ$ about $\ehat$, we get
\begin{align}
    B^{n_1n_2n_3}_{\bk_1,\bk_2,\bk_3} &= - (2\pi)^9\ci f\cH\frac{\mu_3}{k_3}\int_{\{\bq_a\}} F_{n_1}(\{\bq_\a\})F_{n_2}(\{\bq_\b\})G_{n_3}(\{\bq_\g\})\avg{\prod_{a=1}^{n_{123}} \de^{(1)}(-{\sf R}\bq_a)}'_{\rm PV}\nn \\
    &\hspace{6em} \times \de^D\lr({\bk_1 - \sum_{a\in\a}\bq_a})\de^D\lr({\bk_2 - \sum_{a\in\b}\bq_a})\de^D\lr({\bk_3 - \sum_{a\in\g}\bq_a}) \nn \\
    &= - (2\pi)^9\ci f\cH\frac{\mu_3}{k_3}\int_{\{\bq_a\}} F_{n_1}(\{ -{\sf R}^{-1}\bq_\a\})F_{n_2}(\{ -{\sf R}^{-1}\bq_\b\})G_{n_3}(\{ -{\sf R}^{-1}\bq_\g\})\avg{\prod_{a=1}^{n_{123}} \de^{(1)}(\bq_a)}'_{\rm PV}\nn \\
    &\hspace{6em} \times \de^D\lr({\bk_1 +  {\sf R}^{-1} \sum_{a\in\a}\bq_a})\de^D\lr({\bk_2 +  {\sf R}^{-1} \sum_{a\in\b}\bq_a})\de^D\lr({\bk_3 +  {\sf R}^{-1} \sum_{a\in\g}\bq_a}) \;,
\end{align}
where in the second step we have renamed all dummy variables $\bq_a \to \bq_a' = -{\sf R}\bq_a$. Note that the integration measure (and domain) is invariant under this transformation. Since the $180^\circ$ rotation is performed in the plane spanned by the external momenta, we have $\bk_i = -{\sf R}^{-1}\bk_i$. The Dirac delta function can thus be written as 
\begin{align}
    \de^D\lr({\bk_i +  {\sf R}^{-1} \sum_{a}\bq_a}) = \de^D\lr({-{\sf R}^{-1}\lr({\bk_i -\sum_a \bq_a }) }) = \de^D\lr({\bk_i - \sum_{a\in\a}\bq_a})\;,
\end{align}
in which the last equality stems from the relation $\de^D({\sf A}\bk) = \de^D(\bk)/|\det {\sf A} |$, where ${\sf A}$ is a matrix. In this case, the rotation matrices have unit determinant. Joint with the fact that the SPT kernels are invariant under simultaneous rotation and flipping of all arguments, we get
\begin{align}
    B_{\bk_1,\bk_2,\bk_3}^{n_1 n_2 n_3} = -B_{\bk_1,\bk_2,\bk_3}^{n_1 n_2 n_3} = 0\;,
\end{align}
and hence PVICs give vanishing contributions to $\avg{\de\de v_\parallel}$ at all orders in SPT. 

\section{Toy model: scaling universes}\label{sec:scale}
In the following, we show an explicit calculation for the parity-violating contribution to \ddm in  the plane-parallel limit for scaling universes \cite{Scoccimarro:1995if,Scoccimarro:1996se,Pajer:2013jj,Carrasco:2013sva,Carrasco:2013mua}, i.e.\ cosmologies with a power-law matter power spectrum $\plin(k) = \cA k^{n}$. For the ultra-squeezed template \eqref{eq:will-temp}, the general form is given by 
\begin{align}
    \avg{\de(\bk_1)\de(\bk_2)p_\parallel(\bk_3)}' = \cA^3f\cH\sum_{\s\in S_4}\sign(\s) k_1^{n_{\s(1)} }k_2^{n_{\s(2)}} \int_\bq (\nhat\vdot\qhat)(\qhat\vdot(\khat_1\times\khat_2))q^{n_{\s(3)}}|\bk_3-\bq|^{n_{\s(4)}}\;,
\end{align}
where we leave the indices arbitrary in this calculation. Since $|\bk_3-\bq| = k_3^2+q^2-2k_3q \mu$, where $\mu = \khat_3\vdot\qhat$, the angular integral can be separated by taking $\khat_3$ as the polar axis. The azimuthal integral can then be done to give
\begin{align}
    \int \dd\phi_{\bk_3}~(\hat{n}_i \hat{q}^i)(\eps_{abc}\hat{k}_1^a\hat{k}_2^b\hat{q}^c) &= \hat{n}_i (\eps_{abc}\hat{k}_1^a\hat{k}_2^b)\lr[{ \pi(1-\mu^2)\de^{ic} +(2\pi\mu^2-\pi(1-\mu^2))\hat{k}_3^i \hat{k}_3^c }]\\
    &=\pi(1-\mu^2)(\nhat\vdot(\khat_1\times\khat_2))\;,
\end{align}
where the second term containing $\khat_3\times(\khat_1\vdot\khat_2)$ vanishes by momentum conservation. This is exactly the tensorial structure we expected in \eqref{eq:ddp-pv}. With the scalar triple product now independent of the loop momentum $\bq$, we can recast the residual quantities back into an integral over $\dd^3\bq$ by dividing over $2\pi$ and write
\begin{align}\label{eq:to-be-dim-reg}
    \avg{\de(\bk_1)\de(\bk_2)p_\parallel(\bk_3)}' =  \nhat\vdot(\khat_1\times\khat_2)\frac{\cA^3f\cH }{2}\sum_{\s\in S_4}\sign(\s) k_1^{n_{\s(1)} }k_2^{n_{\s(2)}} \int_\bq ({1-(\khat_3\vdot\qhat)^2})q^{n_{\s(3)}}|\bk_3-\bq|^{n_{\s(4)}}\;.
\end{align}
The integral in \eqref{eq:to-be-dim-reg} is linked to the standard one-loop massless propagator in quantum field theory, and is widely used to calculate loop diagrams in LSS \cite{Scoccimarro:1996jy,Pajer:2013jj,Simonovic:2017mhp}. It can be calculated using Feynman parameters, with the result
\begin{align}\label{eq:massless-1L}
    \int \frac{\dd^d\bq}{q^{2\nu_1}|\bk-\bq|^{2\nu_2}}  = \pi^{d/2}k^{d-2\nu_1-2\nu_2}\lr[{\frac{\G\lr({\frac{d}{2}-\nu_1})\G\lr({\frac{d}{2}-\nu_2})\G\lr({\nu_1+\nu_2-\frac{d}{2}})}{\G(\nu_1)\G(\nu_2)\G(d-\nu_1-\nu_2)} }]\;,
\end{align}
where we have provided the full $d$-dimensional result here which would be required to extract the logarithmic divergences if they occur, at which point dimensional regularisation (and renormalisation) has to be performed. Nevertheless, there exist choices of $\nu_{1,2}$ for which \eqref{eq:massless-1L} is convergent, hence we only quote the finite contributions hereinafter. This gives \eqref{eq:to-be-dim-reg} to be 
\begin{align}\label{eq:ddp-power-law}
    \avg{\de(\bk_1)\de(\bk_2)p_\parallel(\bk_3)}' =  \nhat\vdot(\khat_1\times\khat_2)\frac{\cA^3f\cH }{4\pi^{3}}\sum_{\s\in S_4}\sign(\s) k_1^{n_{\s(1)} }k_2^{n_{\s(2)}} \cI(n_{\s(3)},n_{\s(4)})k_3^{3+n_{\s(3)}+n_{\s(4)}}\;,
\end{align}
with the function
\begin{align}\label{eq:Imn_gs}
    \cI(a,b) &= \frac{\G(2+a)\G(-4-a-b)\G(4+b)}{a(2+b)(6+a+b)}\lr[{\sin(a\pi)+\sin(b\pi) - \sin\lr({(a+b)\pi})}]\;.
\end{align}
Since $\cI(a,b)$ is not symmetrical in its two arguments, \ddm does not vanish trivially after summing over permutations in \eqref{eq:ddp-power-law}, provided that its arguments are not both integer-valued. As such, for generic values of power law indices, this gives a concrete example of a non-vanishing parity-violating signal that \ddm is sensitive to. 

\section{Angle-averaged bispectrum}\label{sec:ang-avg}
In this appendix, we derive the angle-averaged bispectrum for the bispectrum $B^{\de\de\De}_{\bell\bfm}$ discussed in section \ref{sec:harmonic-analysis}, for the general form of the parity-violating contribution \eqref{eq:ddp-pv}. We write the harmonic components in terms of the Fourier space bispectrum
\begin{align}
    B^{\de\de\De}_{\bell\bfm} &= \int_{\chi_1,\chi_2,\chi_3} \hspace{-2em} W_\de(\chi_1)W_\de(\chi_2)W_\De(\chi_3)\int \dd^2\nhat_1\;\dd^2\nhat_2\;\dd^2\nhat_3~ Y^*_{\ell_1m_1}(\nhat_1)Y^*_{\ell_2m_2}(\nhat_2)Y^*_{\ell_3m_3}(\nhat_3)\nn \\
    & \hspace{2em} \times \int_{\{\bk_a\}} \ce^{-\ci\sum_a\bk_a\vdot\nhat_a\chi_a} (\nhat_3\vdot(\khat_1\times\khat_2)) (2\pi)^3\de^D(\bk_{123})A_\mathsmaller{\rm PV}(k_1,k_2,k_3)\;,
\end{align}
where the LoS is now a local entity attached to the $\De$ field. We have also absorbed the redshift-dependent factors $f(\chi)\cH(\chi)$ into the window function $W_\De(\chi)$. As before, we can expand the scalar triple product in terms of spherical harmonics 
\begin{align}
    \nhat_3\vdot(\khat_1\times\khat_2) = -\ci(2\pi)^{3/2} \frac{\sqrt{2}}{3} \sum_{M_1, M_2, M_3}\begin{pmatrix}
        1 & 1 & 1 \\
        M_1 & M_2 & M_3
    \end{pmatrix}
   Y_{1M_1}(\khat_1)Y_{1M_2}(\khat_2) Y_{1M_3}(\nhat_3)\;.
\end{align}
After performing the Rayleigh expansion, the $\nhat_{1,2}$ integrals can be done with orthogonality relations, whereas the $\nhat_3$ integral gives the Gaunt symbol \eqref{eq:gaunt}. Altogether, it reads
\begin{align}
    B^{\de\de\De}_{\bell\bfm} &= -(4\pi)^{9/2}\frac{\sqrt{2}}{3}\ci^{-\ell_{12}}\sum_{L,M}\sum_{M_1,M_2,M_3} (-1)^{m_3}\ci^{-L+1}
    \begin{pmatrix}
        1 & 1 & 1 \\
        M_1 & M_2 & M_3
    \end{pmatrix}
    \cG_{1L\ell_3}^{M_3M-m_3}\nn \\
    &\hspace{2em}\times \int_{\chi_1,\chi_2,\chi_3 }\hspace{-2em} W_\de(\chi_1)W_\de(\chi_2)W_\De(\chi_3)\int_{\{\bk_a\}}j_{\ell_1}(k_1\chi_1)j_{\ell_2}(k_2\chi_2)j_L(k_3\chi_3)\nn\\
    &\hspace{2em}\times Y^*_{\ell_1m_1}(\khat_1)Y_{1M_1}(\khat_1) Y^*_{\ell_2m_2}(\khat_2)Y_{1M_2}(\khat_2)Y^*_{LM}(\khat_3)(2\pi)^3\de^D(\bk_{123})A_\mathsmaller{\rm PV}(k_1,k_2,k_3)
\end{align}
To proceed, we can do a decomposition of the Dirac delta function, i.e.\ 
\begin{align}\label{eq:dirac-decompose}
    (2\pi)^3\de^D(\bk_{1\ldots n}) = \int \dd^3\bx~ \ce^{\ci\bk_{1\ldots n}\vdot\bx} = \int \dd^3\bx \lr[{\prod_{a=1}^n \sum_{\l_a,\mu_a}4\pi \ci^a j_{\l_a}(k_ax)Y_{\l_a\mu_a}(\khat_a)Y^*_{\l_a\mu_a}(\xhat_a) }]\;,
\end{align}
in which the angular integrals for the momenta $\khat_a$ and the auxiliary variable $\xhat$ can be done via orthogonality relations or \eqref{eq:gaunt}. The result is
\begin{align}\label{eq:before-w9}
    B^{\de\de\De}_{\bell\bfm} &= -(4\pi)^{15/2}\frac{\sqrt{2}}{3}\ci^{-\ell_{12}}\sum_{\l_a,\mu_a,M_a} (-1)^{m_{123}}\ci^{\l_{12}+1}
    \begin{pmatrix}
        1 & 1 & 1 \\
        M_1 & M_2 & M_3
    \end{pmatrix}
    \cG_{1\l_3\ell_3}^{M_3\mu_3-m_3}\cG_{1\l_1\ell_1}^{M_1\mu_1-m_1}\cG_{1\l_2\ell_2}^{M_2\mu_2-m_2}\cG_{\l_1\l_2\l_3}^{\mu_1\mu_2\mu_3}\nn \\
    &\hspace{2em}\times \int_{\chi_1,\chi_2,\chi_3 }\hspace{-2em} W_\de(\chi_1)W_\de(\chi_2)W_\De(\chi_3)\int \lr[{\prod_{a=1}^3 \frac{k_a^2\dd k_a}{(2\pi)^3}}]j_{\ell_1}(k_1\chi_1)j_{\ell_2}(k_2\chi_2)j_{\l_3}(k_3\chi_3)\nn\\
    &\hspace{2em}\times \int x^2\dd x ~ j_{\l_1}(k_1x)j_{\l_2}(k_2x)j_{\l_3}(k_3x)A_\mathsmaller{\rm PV}(k_1,k_2,k_3)\;.
\end{align}
Writing the Gaunt symbol in terms of the 3j symbols with which it is defined, as in \eqref{eq:gaunt}, the sum over azimuthal numbers $M_a$ and $\mu_a$ can be done with the definition of the Wigner 9j symbol, i.e.
\begin{align}
    \begin{pmatrix}
        \ell_{13} & \ell_{24} & L \\
        m_{13} & m_{24} & M 
    \end{pmatrix}
    \begin{Bmatrix}
        \ell_1 & \ell_2 & \ell_{12} \\
        \ell_3 & \ell_4 & \ell_{34} \\
        \ell_{13} & \ell_{24} & L 
    \end{Bmatrix} = \sum_{\substack{m_1,m_2,m_3,m_4 \\ m_{12},m_{34}}} &
    \begin{pmatrix}
        \ell_1 & \ell_2 & \ell_{12} \\
        m_1 & m_2 & m_{12} 
    \end{pmatrix}
    \begin{pmatrix}
        \ell_3 & \ell_4 & \ell_{34} \\
        m_3 & m_4 & m_{34} 
    \end{pmatrix} \nn\\
    & \hspace{-1em} \times
    \begin{pmatrix}
        \ell_1 & \ell_3 & \ell_{13} \\
        m_1 & m_2 & m_{12} 
    \end{pmatrix}
    \begin{pmatrix}
        \ell_2 & \ell_4 & \ell_{24} \\
        m_2 & m_4 & m_{34} 
    \end{pmatrix}
    \begin{pmatrix}
        \ell_{12} & \ell_{34} & L \\
        m_{12} & m_{34} & M 
    \end{pmatrix}
\end{align}
to get the bispectrum into the form of \eqref{eq:red-bis-def}, 
\begin{align}
    B_{\bell\bfm}^{\de\de\De}= (4\pi)^{15/2}\frac{\sqrt{2}}{3}\ci^{-\ell_{12}}
     \begin{pmatrix}
        \ell_1 & \ell_2 & \ell_3 \\
        m_1 & m_2 & m_3
    \end{pmatrix}
     \sum_{\l_a}\ci^{\l_{12}+1}
    \cQ_{\l_1\l_2\l_3}\cJ^{\ell_1\ell_2}_{\l_1\l_2\l_3}\;,
\end{align}
in which the angle-averaged bispectrum $b^{\de\de\De}_{\bell}$ can be identified. Here, we have written the last two lines of \eqref{eq:before-w9} into $\cJ^{\ell_1\ell_2}_{\l_1\l_2\l_3}$, which encodes the projection effects and shape information among other factors. The angular coupling is encoded in the symbol
\begin{align}\label{eq:Q-ang-coup}
    \cQ_{\l_1\l_2\l_3} = h_{1\l_1\ell_1}h_{1\l_2\ell_2}h_{1\l_3\ell_3}h_{\l_1\l_2\l_3}
    \begin{Bmatrix}
         1 & 1 & 1 \\
         \l_1 & \l_2 & \l_3 \\
         \ell_1 & \ell_2 & \ell_3
     \end{Bmatrix}\;,
\end{align}
in which we have defined
\begin{align}\label{eq:h-symb}
    h_{\ell_1\ell_2\ell_3} = \sqrt{\frac{(2\ell_1+1)(2\ell_2+1)(2\ell_3+1)}{4\pi}}
    \begin{pmatrix}
        \ell_1 & \ell_2 & \ell_3 \\
        0 & 0 & 0 
    \end{pmatrix}
    \;.
\end{align}
Through \eqref{eq:Q-ang-coup} and \eqref{eq:h-symb}, one can read off the selection rules 
\begin{subequations}
\begin{align}
    1+\l_a + \ell_a & \in 2\mathbb{N}\quad {\rm for}~a=1,2,3 \\
    \l_1 + \l_2 + \l_3 & \in 2\mathbb{N}
\end{align}
\end{subequations}
which combined gives $\ell_{123} = {\rm odd}$, which was what we expected from the analysis in section \ref{sec:sphere-symm}.

\section{Efficient evaluation}\label{sec:effeval}
To estimate the Fisher information for the parity-violating amplitude \eqref{eq:fish}, one requires access to the response of the observable to the parity-violating trispectrum $\ptl B_{\bell,\bfm}/\ptl\pnl$. One path to acquire this is through parity-violating simulations of large-scale structure and its effect on the CMB. In this work, we have taken the cost-efficient course and derive analytical expressions for the Fisher information. In this appendix, we present the derivation of the Fisher information, and show that it can be cast in a factorisable form. We also detail the numerical implementation that allows for a fast and efficient computation.
\subsection{Derivation of factorisable Fisher}\label{sec:fact-fish}
In contrast to \eqref{eq:dv||-lm}, where our choice of decomposition parametrises the primordial shape by its diagonal, we adopt the symmetric decomposition
\begin{align}
    \De_{\ell m} = \int \dd^2\nhat~Y^*_{\ell m}(\nhat) \int_0^\infty \dd\chi~W_\De(\chi)\int_{\bk_1,\bk_2}\ce^{-\ci\bk_{12}\vdot\nhat\chi}\lr[{\frac{\ci\nhat\vdot\khat_2}{k_2}}]\de^{(1)}(\bk_1)\de^{(1)}(\bk_2)\;.
\end{align}
Writing the $\nhat\vdot\khat_1$ term as a derivative, and performing the Rayleigh expansion, we get
\begin{align}
    \De_{\ell m} &= -(4\pi)^2\sum_{L_1,M_1}\sum_{L_2,M_2}\ci^{-L_{12}}\int \dd^2\nhat~Y^*_{\ell m}(\nhat) Y_{L_1M_1}(\nhat)Y_{L_2M_2}(\nhat)\int_0^\infty \dd\chi~W_\De(\chi)\nn\\
    &\hspace{2em}\times\int_{\bk_1,\bk_2}\frac{1}{k_2^2}\, j_{L_1}(k_1\chi)\ptl_\chi j_{L_2}(k_2\chi) Y^*_{L_1M_1}(\khat_1)Y^*_{L_2M_2}(\khat_2)\de^{(1)}(\bk_1)\de^{(1)}(\bk_2)
\end{align}
and the $\nhat$ integral can be done to give a Gaunt symbol using \eqref{eq:gaunt}. Joint with the density fields \eqref{eq:de-lm}, we have 
in terms of the curvature fluctuations
\begin{align}
    B_{\ell_1\ell_2L}^{m_1 m_2M} &= (4\pi)^4 \sum_{\ell_3,\ell_4}\sum_{m_3,m_4} \ci^{-\ell_{1234}}(-1)^{M}\cG_{\ell_3\ell_4L}^{m_3 m_4 -M}\int_{\chi_1,\chi_2,\chi} W_\de(\chi_1)W_\de(\chi_2)W_\De(\chi) \nn\\
    & \hspace{2em} \times  \int_{\{\bk_a\}} j_{\ell_1}(k_1\chi_1)j_{\ell_2}(k_2\chi_2)j_{\ell_3}(k_3\chi)\ptl_\chi j_{\ell_4}(k_4\chi) Y_{\ell_1m_1}^*(\khat_1) Y_{\ell_2m_2}^*(\khat_2) Y_{\ell_3m_3}^*(\khat_3) Y_{\ell_4m_4}^*(\khat_4) \nn \\
    &\hspace{2em} \times \cM_\de(k_1) \cM_\de(k_2) \cM_\de(k_3) \cM_v(k_4) \avg{\z({\bk_1})\z({\bk_2})\z({\bk_3})\z({\bk_4})}\;,
\end{align}
in which we have used the multipole convention in section \ref{sec:estimator}, i.e.\ $B_{\ell_1\ell_2L}^{m_1 m_2M} \equiv \avg{\de_{\ell_1 m_1}\de_{\ell_2 m_2}\De_{LM}}$. The transfer functions $\cM_{\de/v}(k)$ are given in \eqref{eq:transfer}. For any term in the ultra-squeezed template \eqref{eq:will-temp}, we can decompose the delta function as in \eqref{eq:dirac-decompose} and write it as
\begin{align}
    \avg{\z({\bk_1})\z({\bk_2})\z({\bk_3})\z({\bk_4})} &= \ci  \frac{9}{25}\pnl A^3(2\pi)^3\de^D(\bk_{1234})(\bk_1\vdot(\bk_2\times\bk_3))k_1^{n_1}  k_2^{n_2} k_3^{n_3} k_4^{n_4}\\ 
    &= - \frac{9}{25}\pnl A^3k_1^{n_1} k_2^{n_2} k_3^{n_3} k_4^{n_4}\int_{\bx}\epsilon^{abc}\,\lr({\grad_a \ce^{\ci\bk_1\vdot\bx}})\lr({\grad_b \ce^{\ci\bk_2\vdot\bx}})\lr({\grad_c \ce^{\ci\bk_3\vdot\bx}})\ce^{\ci\bk_4\vdot\bx}\;.
\end{align}
 Note that the indices $n_i$'s here differ from that in \eqref{eq:will-temp} by $-3$, as we have included the scaling in the primordial power spectrum $P_\z(k)$ in \eqref{eq:pv-zeta}. Given that the Rayleigh expansion provides a natural separation between the radial and angular parts, the scalar triple product is best written in the basis $\{\ptl_x, \up, \down\}$, where $\up$ and $\down$ are the Newman-Penrose spin operators, i.e.
\begin{align}
    \nabla f(\bx)\vdot(\nabla g(\bx) \times \nabla h(\bx)) = \frac{\ci}{2x^2}\lr[{\ptl_x f\lr({\up g\down h - \down g \up h})}] + {\rm cyclic\ permutations}\;.
\end{align}
Using the orthogonality relations for the angular integrals $\khat_a$, we get 
\begin{align}
    B_{\ell_1\ell_2 L}^{m_1m_2M} &= -\ci \frac{9}{25}\pnl A^3\sum_{\tt ijk}\sum_{\tt abcd}\int\frac{\dd x}{2}\int_{\xhat} \int_{\chi_1} W_\de(\chi_1)\int_{\chi_2} W_\de(\chi_2) \int_\chi W_\De(\chi) \nn\\
     &\hspace{2em}\times \eps_{\tt ijk}\eps_{\mathtt{abcd}}{\color{Blue3}\Big[{{^\mathtt{i}}G^\mathtt{a}_{\ell_1}(x,\chi_1) (-1)^{m_1} {_\mathtt{i}Y_{\ell_1 -m_1}}(\xhat)}\Big]}{\color{Red}\Big[{ {^\mathtt{j}}G^\mathtt{b}_{\ell_2}(x,\chi_2) (-1)^{m_2} {_\mathtt{j}Y_{\ell_2 -m_2}}(\xhat)}\Big]} \int_{\nhat}Y^*_{LM}(\nhat)\nn\\
    &\hspace{2em}\times 
    {\color{orange} \lr[{ \sum_{\ell_3,m_3} {^\mathtt{k}}G_{\ell_3}^\mathtt{c}(x,\chi)(-1)^{m_3} {_\mathtt{k}Y_{\ell_3-m_3}}(\xhat) Y_{\ell_3 m_3}(\nhat) }]}
    {\color{Green2} \lr[{ \sum_{\ell_4,m_4} V^\mathtt{d}_{\ell_4}(x,\chi)(-1)^{m_4} {Y_{\ell_4-m_4}}(\xhat) Y_{\ell_4 m_4}(\nhat) }]}\;.
\end{align}
which can be identified with \eqref{eq:B_ggT} with appropriate window functions. The kernels here are defined in \eqref{eq:GV-kernels}. With this, the Fisher information \eqref{eq:fish} can then be written as 
\begin{align}\label{eq:B^2}
    F &=\frac{A^6}{8}\lr({\frac{9}{25}})^2\sum_{\mathtt{ijk}}\sum_{\mathtt{abcd}}\sum_{\mathtt{i'j'k'}}\sum_{\mathtt{a'b'c'd'}} \eps_{\mathtt{ijk}}\eps_{\mathtt{abcd}} \eps_{\mathtt{i'j'k'}}\eps_{\mathtt{a'b'c'd'}}\nn\\
    &\hspace{2em}\times \int{\dd x}\int{\dd x'}\int_\chi W_\De(\chi) \int_{\chi'} W_\De(\chi') \int_{\xhat}\int_{\xhat'}\int_{\nhat}\int_{\nhat'} {\color{Brown} \lr[{\sum_{L,M} \frac{Y^*_{LM}(\nhat)Y_{LM}(\nhat')}{C_\ell^{TT} } }] } \nn\\
    &\hspace{2em}\times {\color{Blue3} \lr[{ \sum_{\ell_1,m_1} \frac{ {^\mathtt{i}}G^\mathtt{a}_{\ell_1} (x){^\mathtt{i'}}G^\mathtt{a'}_{\ell_1}(x')}{C_{\ell_1}^{gg}}  {_\mathtt{i}}Y_{\ell_1 m_1}(\xhat)  {_\mathtt{i'}}Y^*_{\ell_1m_1}(\xhat') }] }  {\color{Red}\lr[{ \sum_{\ell_2,m_2} \frac{{^\mathtt{j}}G_{\ell_2}^\mathtt{b}(x){^\mathtt{j'}}G_{\ell_2}^\mathtt{b'}(x')}{C_{\ell_2}^{gg}} {_\mathtt{j}}Y_{\ell_2 m_2}(\xhat)  {_\mathtt{j'}}Y^*_{\ell_2m_2}(\xhat') }]  }\nn\\
    &\hspace{2em}\times {\color{orange} \lr[{ \sum_{\ell_3,m_3}{{^\mathtt{k}}G_{\ell_3}^\mathtt{c}(x,\chi)}  {_\mathtt{k}Y_{\ell_3m_3}}(\xhat) Y^*_{\ell_3 m_3}(\nhat) }]  }{\color{Green2}\lr[{ \sum_{\ell_4,m_4}{V^\mathtt{d}_{\ell_4}(x,\chi)}  {Y_{\ell_4m_4}}(\xhat) Y^*_{\ell_4 m_4}(\nhat) }]   } \nn\\
    &\hspace{2em}\times {\color{orange} \lr[{ \sum_{\ell_3',m_3'}{{^\mathtt{k'}}G^\mathtt{c'}_{\ell_3'}(x',\chi')}  {_\mathtt{k'}Y^*_{\ell_3'm_3'}}(\xhat') Y_{\ell_3' m_3'}(\nhat') }] }{\color{Green2} \lr[{ \sum_{\ell_4',m_4'}{V^\mathtt{d'}_{\ell_4'}(x',\chi')}  {Y^*_{\ell_4'm_4'}}(\xhat') Y_{\ell_4' m_4'}(\nhat') }]   }\;. 
\end{align}
The sum over $m_1$ and $m_2$ in the third line of \eqref{eq:B^2} can be done by applying \eqref{eq:swsh-prod} to spin-weighted spherical harmonics pairs at coincident points and using the orthogonality relation \eqref{eq:ortho-m}, i.e.
\begin{align}
    &\sum_{m_1,m_2}\lr[{{\color{Blue3}{_{\tt i}}Y_{\ell_1m_1}(\xhat)}{\color{Red}{_{\tt j}}Y_{\ell_2m_2}(\xhat)}}]\lr[{{\color{Blue3}{_{\tt i'}}Y^*_{\ell_1m_1}(\xhat')}{\color{Red}_{\tt j'}Y^*_{\ell_2m_2}(\xhat')}}] \\
    = ~& \sum_{m_1,m_2} \lr[{ \sum_{L,M}\L_{\ell_1\ell_2L}
    \begin{pmatrix}
        \ell_1 & \ell_2 & L \\
        m_1 & m_2 & M 
    \end{pmatrix} 
    \begin{pmatrix}
        \ell_1 & \ell_2 & L \\
        -{\tt i} & -{\tt j} & {\tt i+j} 
    \end{pmatrix} {_{\tt i+j}}Y^*_{LM}(\xhat)
    }]\nn\\
    &\hspace{1.5em}\times\lr[{ \sum_{L',M'}\L_{\ell_1\ell_2L'}
    \begin{pmatrix}
        \ell_1 & \ell_2 & L' \\
        m_1 & m_2 & M' 
    \end{pmatrix} 
    \begin{pmatrix}
        \ell_1 & \ell_2 & L' \\
        -{\tt i'} & -{\tt j'} & {\tt i'+j'} 
    \end{pmatrix} {_{\tt i'+j'}}Y_{L'M'}(\xhat')
    }]\\
    =~ &\sum_{L,M} \frac{\L_{\ell_1\ell_2L}^2}{2L+1} \lr[{\begin{pmatrix}
        \ell_1 & \ell_2 & L \\
        -{\tt i} & -{\tt j} & {\tt i+j} 
    \end{pmatrix}\begin{pmatrix}
        \ell_1 & \ell_2 & L \\
        -{\tt i'} & -{\tt j'} & {\tt i'+j'} 
    \end{pmatrix}}]
    {_{\tt i+j}}Y^*_{LM}(\xhat){_{\tt i'+j'}}Y_{LM}(\xhat')\\
    =~ &\sum_{L,M} \frac{\L_{\ell_1\ell_2L}^2}{2L+1} \lr[{\int_{-1}^1 \dd(\cos\th) ~d^{\ell_1}_{\tt -i,-i'}(\th)d^{\ell_2}_{\tt -j,-j'}(\th)d^{L}_{\tt i+j,i'+j'}(\th) }]
    {_{\tt i+j}}Y^*_{LM}(\xhat){_{\tt i'+j'}}Y_{LM}(\xhat')
\end{align}
where we have rewritten the Wigner 3j pairs into its integral form \eqref{eq:triple-d}, and denoted $\L_{\ell_1\ell_2\ell_3} = \sqrt{(2\ell_1+1)(2\ell_2+1)(2\ell_3+1)/4\pi}$ for brevity. Performing the same calculation for the fourth and fifth line of \eqref{eq:B^2}, there appears exactly two spherical harmonics for each unit vector field $\nhat,\nhat',\xhat,$ and $\xhat'$, allowing all four angular integrals to be done using orthogonality relations. This gives
\begin{align}
    F &=\frac{A^6}{8}\lr({\frac{9}{25}})^2 \int{\dd x}\int{\dd x'}\sum_{\tt ijk}\sum_{\tt i'j'k'}\sum_{\tt abcd}\sum_{\tt a'b'c'd'}  \eps_{\tt ijk} \eps_{\tt i'j'k'}\eps_{\tt abcd} \eps_{\tt a'b'c'd'} \nn\\
    &\hspace{7em}\times \sum_L \frac{2L+1}{(4\pi)^3} \frac{1}{C_L^{TT}}{\color{Teal} G^\mathtt{aa'bb'}_{{\tt ii'jj'},L}(x,x')}{\color{magenta}X^\mathtt{cd}_{{\tt k},L}(x)}{\color{magenta}X^\mathtt{c'd'}_{{\tt k'},L}(x')}\;,
\end{align}
where the kernels are defined as
\begin{subequations}
\begin{align}
    {\color{Teal} G^\mathtt{aa'bb'}_{{\tt ii'jj'},L}(x,x')} &= \frac{1}{2}\int_{-1}^1 \dd\cos\th\  d^L_{\tt i+j,i'+j'}(\th) {\color{Blue3}\lr[{ \sum_{\ell_1} \frac{ {^\mathtt{i}}G^\mathtt{a}_{\ell_1} (x){^\mathtt{i'}}G^\mathtt{a'}_{\ell_1}(x')}{C_{\ell_1}^{gg}}  (2\ell_1+1) d^{\ell_1}_{\mathtt{i'i}}(\theta) }]} \nn\\
    &\hspace{11em}\times{\color{Red}\lr[{ \sum_{\ell_2} \frac{ {^\mathtt{j}}G^\mathtt{b}_{\ell_1} (x){^\mathtt{j'}}G^\mathtt{b'}_{\ell_1}(x')}{C_{\ell_2}^{gg}}  (2\ell_2+1)d^{\ell_2}_{\mathtt{j'j}}(\theta) }] }\;,\\
    {\color{magenta}X^\mathtt{cd}_{{\tt k},L}(x)}&= \frac{1}{2}\int_{-1}^1 \dd\cos\th ~d^L_{\tt k0}(\th) \int \dd\chi~ W_\De(\chi) {\color{orange}\lr[{\sum_{\ell_3} {^{\tt k}G^\mathtt{c}_{\ell_3}(x,\chi)}  (2\ell_3+1 ) d^{\ell_3}_{\tt 0k}(\th) }]  } \nn\\
    &\hspace{14.5em}\times{\color{Green2} \lr[{ \sum_{\ell_4}{V^\mathtt{d}_{\ell_4}(x,\chi)}  (2\ell_4+1 )d^{\ell_4}_{\tt 00}(\th)  }] }\;.\label{eq:Xcd}
\end{align}
\end{subequations}
There are many symmetries that can be exploited to simplify the numerical evaluation of the Fisher matrix. Mainly, the combination  
\begin{align}
    H_{{\tt kk'},L}  \equiv \int\dd x\int\dd x' \sum_{\tt ij}\sum_{\tt i'j'}\sum_{\tt abcd}\sum_{\tt a'b'c'd'}  \eps_{\tt ijk} \eps_{\tt i'j'k'}\eps_{\tt abcd} \eps_{\tt a'b'c'd'} {\color{Teal} G^\mathtt{aa'bb'}_{{\tt ii'jj'},L}}{\color{magenta}X^\mathtt{cd}_{{\tt k},L}}{\color{magenta}X^\mathtt{c'd'}_{{\tt k'},L}}\;,
\end{align}
obeys $H_{{\tt kk'},L} = H_{{\tt k'k},L}$. Recognising that ${\tt i+j+k} = 0$, the symmetries of the Wigner $d$-matrices \eqref{eq:wig-d-symms} also give $H_{{\tt kk'},L} = H_{{\tt -k-k'},L}$. The calculation of the Fisher information hence boils down to evaluating
\begin{align}\label{eq:F-decomposed}
    F = \frac{A^6}{8}\lr({\frac{9}{25}})^2\sum_L \frac{2L+1}{(4\pi)^3} \frac{1}{C_L^{TT}}\lr({H_{{\tt 00},L} + 4H_{{\tt 01},L} + 2H_{{\tt 11},L}+ 2H_{{\tt 1-1},L}})\;.
\end{align}

\begin{figure}
    \centering
    \includegraphics[width=\linewidth]{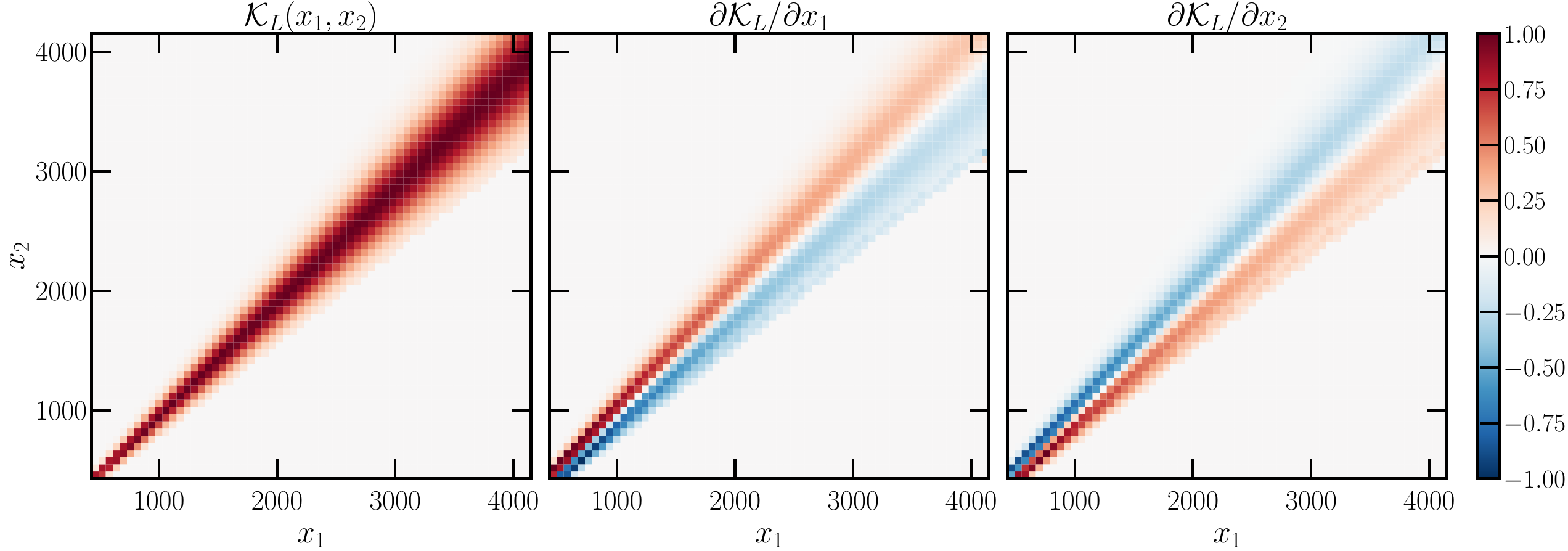}
    \caption{Structure of the integral $\cK_L(x_1,x_2)$ \eqref{eq:Kx1x2-int} and its first derivatives for $L=50$. The plots have been normalised to their corresponding maximum values for easy comparison.}
    \label{fig:kx1x2}
\end{figure}

\begin{figure}
    \centering
    \includegraphics[width=0.7\linewidth]{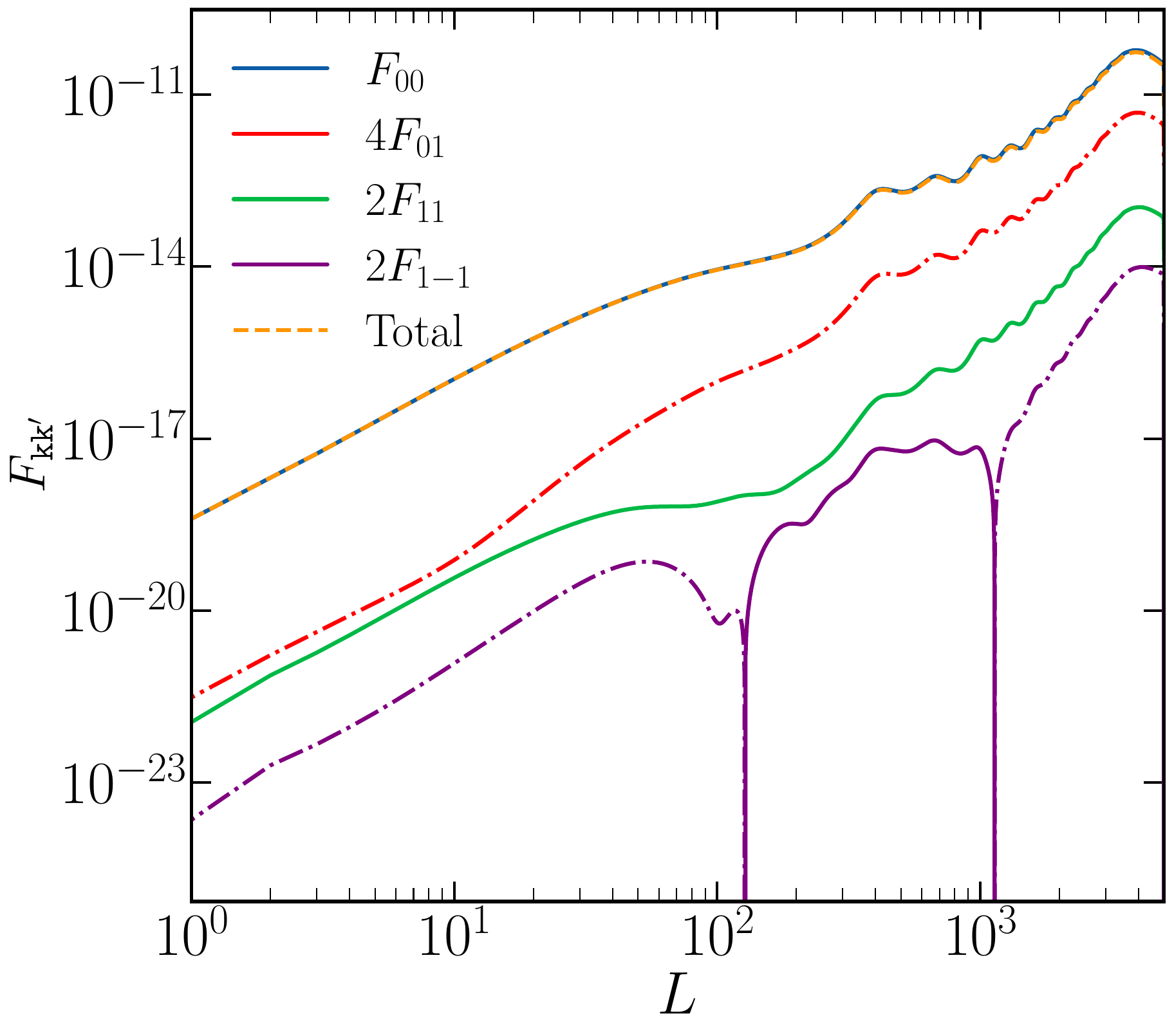}
    \caption{Contributions of the individual components to the Fisher information, where $F_\mathtt{kk'}$'s are $H_\mathtt{kk'}$'s in \eqref{eq:F-decomposed} with prefactors multiplied. Solid and dotted-dashed lines represent positive and negative values respectively. The yellow dashed line is the co-added Fisher information $F_L$, which largely coincides with $F_{\mathtt{00},L}$.}
    \label{fig:ind-fish}
\end{figure}

Before we show the numerical result, it helps to gain intuition about the relative importance of the terms appearing in \eqref{eq:F-decomposed}. A hint is that the kernels \eqref{eq:GV-kernels} contain the integral
\begin{align}\label{eq:Kx1x2-int}
    \cK_L(x_1,x_2) =  4\pi \int_0^\infty \dd \ln k\ k^n j_L(kx_1)j_L(kx_2)
\end{align}
and its first derivatives, either with respect to $x_1$ or $x_2$. Since the integral is symmetric under exchange of its arguments, i.e.\ $\cK_L(x_1,x_2)=\cK_L(x_2,x_1)$, its first derivatives necessarily flip sign across the diagonal, displaying a dipolar nature. A visualisation of the integral is shown in figure \ref{fig:kx1x2}. As such, when the derivatives of $\cK_L(x_1,x_2)$ are integrated over $x_1$ or $x_2$ with a broad window function, the integral is largely suppressed. In fact, this is a well-known result in the literature of kSZ velocity reconstruction \cite{Contreras:2022zdz,McCarthy:2024nik,Lai:2025qdw,McCarthy:2025brx,EmbilVillagra:2026fnh} and is the mathematical reason behind the vanishing of velocity-galaxy power spectra in equal redshift bin cross-correlations.
In our case, the parity-violating trispectra modifies the contribution from the kSZ, such that it entails an integral of either $\cK_L \times \ptl\cK_L/\ptl x_2$ or $ \ptl\cK_L/\ptl x_1 \times \ptl\cK_L/\ptl x_2$ in ${\color{magenta}X^\mathtt{cd}_{{\tt k},L}(x_1)}$ \eqref{eq:Xcd}, of which the latter contains large cancellations through its dipolar structure. As such, the contributions to the Fisher information in \eqref{eq:F-decomposed} have a hierarchy
\begin{align}
    H_{{\tt 00},L} \geq |H_{{\tt 01},L}|  \geq H_{{\tt 11},L} \geq H_{{\tt 1-1},L}
\end{align}
at large $L$'s.\footnote{At small $L$'s, the broad support of spherical Bessel functions in \eqref{eq:Kx1x2-int} do not cancel sufficiently in $\cK_L\times \ptl\cK_L/\ptl x$.} The final inequality between $H_{{\tt 11}}$ and $H_{{\tt 1-1}}$ is from the (semi-)positivity of $H_{{\tt 11},L}$. The numerical result of the individual components are shown in figure \ref{fig:ind-fish} which confirms the hierarchy above.\footnote{The scans over shot noise, CMB noise, and $\ell_{\rm min/max}$ in section \ref{sec:results} were done with all contributions to the Fisher information despite the hierarchy. However, future forecasts with more complicated setups could opt to drop these subdominant contributions in exchange for speed.}

\subsection{Numerical implementation}\label{sec:fftlog}
The computational bottleneck of the Fisher information in its factorised form is the kernels \eqref{eq:GV-kernels}, largely owing to the oscillatory nature of the spherical Bessel functions. To proceed, we follow the procedure outlined in \cite{Assassi:2017lea} for fast evaluation of angular statistics (see also \cite{Lee:2020ebj} for application towards angular trispectra and further development of this machinery). As some of our implementations deviate from usual practices, we outline the general background that will be useful to understand these differences, and we point the readers to references above for more details. 

\subsubsection{FFTLog decomposition and Weber-Schafheitlin integrals}\label{sec:ws-integrals}
First, we perform a FFTLog decomposition \cite{Hamilton:1999uv} of the function $\cM_\de(k)$ into an $(N+1)$-dimensional basis
\begin{align}
     \cM_\de(k) = \sum_{m=-N/2}^{N/2} c_m k^{b+\ci\eta_m}\;,\quad \text {where }\eta_m = \frac{2\pi m}{\log(k_{\rm max}/k_{\rm min})}
\end{align}
over the bulk of the support $[k_{\rm min},k_{\rm max}]$ of $\cM_\de(k)$. The free parameter $b$ here is the FFTLog bias which is hand-picked to ensure convergence of the decomposition. For $\cM_\de(k)$, we find $b=0.7$ to give the best performance, with $N=255$ points in the range $k\in[7\times 10^{-6},10^2]\ {\rm Mpc}^{-1}$.

\vskip 5pt
The decomposition above (along with the power-law scaling for the parity-violating trispectrum) allows all the integrals in \eqref{eq:GV-kernels} to be written in the form of $\cK_L(x,\chi)$ \eqref{eq:Kx1x2-int} with appropriate indices $n$. These integrals can be written in terms of the Weber-Schafheitlin (WS) integrals \cite{Watson:1944:TBF,Dixon:1930:IIT}, whose solutions are analytically known to be
\begin{align}\label{eq:Ilnt}
    I_L(n,t)&\equiv 4\pi\int_0^\infty \dd\ln y\ y^n j_L(y)j_L(yt) \nn\\
    &= \frac{2^{n-1}\pi^2\G\lr({L+\frac{n}{2}})}{\G\lr({\frac{3-n}{2}})\G\lr({L+\frac{3}{2}})}\ t^L{_2}F_1\lr({\frac{n-1}{2}, L+\frac{n}{2}, L+\frac{3}{2}\;\Bigg|\; t^2})\equiv \tilde{I}_L(n,t)
\end{align}
 in the region $|t|\leq 1$ where it converges $( {\rm i.e.}\ -2L<\Re(n)<3$ for $t\neq 1$ and $-2L<\Re(n)<2$ at $t=1)$. Here, the function ${_2}F_1$ is the hypergeometric function.  For the region $|t|>1$, the same integral can be computed by a redefinition of variables
\begin{align}\label{eq:reflect}
    I_L(n,t) = t^{-n} I_L(n,t^{-1})\;.
\end{align}

\vskip 5pt
Returning to the convergence properties of the WS integral \eqref{eq:Ilnt}, note that we have purposefully distinguished between its definition $I_L(n,t)$, which is associated with a convergence condition in $n$, to its solution $\tilde{I}_L(n,t)$, which is well-defined for any $n$. 
Outside of the radius of convergence, the WS integral should be formally understood in terms of distributions \cite{Wrochna:2010WSI,Lee:2020ebj}. For example, one can easily identify that the (upper) boundary of the radius of convergence, $n=3$, corresponds to the closure equation for spherical Bessel functions:
\begin{align}
    4\pi\int_0^\infty \dd k\ k^2 j_L(kx_1)j_L(kx_2) = \frac{2\pi^2}{x_1^2} \de^D(x_1-x_2)\;.
\end{align}
For $n\geq 3$, powers of $k$ can be traded for derivatives by recognising that the spherical Bessel functions satisfy a differential equation \cite{Assassi:2017lea,Schoneberg:2018fis}
\begin{align}\label{eq:cov-divs}
    \cD_L(x) j_L(kx)\equiv  \lr[{-\ptl_x^2 - \frac{2}{x} \ptl_x +\frac{L(L+1)}{x^2}}]j_L(kx) = k^2 j_L(kx)\;.
\end{align}
This means that one can obtain the following identity, 
\begin{align}\label{eq:ILdef}
    I_L(n+2M,t) = [\cD_L(t)]^M I_L(n,t)\;,
\end{align}
which holds at the level of distributions, allowing for higher-order WS integrals to be evaluated in terms of lower-order ones. This is vital to compute all the kernels necessary for \eqref{eq:GV-kernels}, which require $n-b \in[-2,-1,0,3]$ for ${^{\tt s}}G^{\tt n}_\ell$ and $n-b \in[-4,-3,-2,1]$ for $V^{\tt n}_\ell$, one of which necessitates the use of \eqref{eq:ILdef} for our choice of FFTLog bias $b=0.7$. 

\vskip 5pt
In computing angular statistics, one typically progresses by performing an integration-by-parts and shifting the derivative operator $\cD_L$ onto window functions in the line-of-sight integrals. Unfortunately, this technique fails for the kSZ kernels \eqref{eq:Xcd} due to the composite nature of the momentum field. Instead, we use the fact that \eqref{eq:cov-divs} is also a valid identity for the solution $\tilde{I}_L$ \cite{Lee:2020ebj}, i.e.
\begin{align}
    \tilde{I}_L(n+2M,t) = [\cD_L(t)]^M \tilde{I}_L(n,t)\;,
\end{align}
for all values of $n$, and evaluate the derivatives in $\cD_L$ numerically on the lower-order WS integral $\tilde{I}_L(n,t)$. In other words, we represent $I_L(n+2M,t)$ as $[\cD_L(t)]^M \tilde{I}_L(n,t)$ in our numerical evaluation.\footnote{Choosing to represent these divergent integrals as operator kernels of the differential operator $\cD_L$ essentially fixes the values of contact terms of $I_L$, which is a reasonable choice given that the spherical Bessel functions originate from a plane-wave expansion via the spherical Laplacian operator.} 

\subsubsection{Consistency check}
To understand if the implementation of divergent integrals through the method outlined above in appendix \ref{sec:ws-integrals} truly give results that represent our system, we want to benchmark products of our implementation that can be verified with other means. With this in mind, recall that the projected galaxy field in harmonic space 
\begin{align}
    g_{\ell m} = 4\pi\ci^{-\ell} b_1 \int\dd\chi\  W_g(\chi) \int_{\bk}j_\ell(k\chi) Y^*_{\ell m}(\khat)\cM_\de(\bk)\z(\bk)\;,
\end{align}
has the power spectrum
\begin{align}\label{eq:clgg-recon}
    C_{\ell}^{gg} = A\int_0^\infty \dd x\ x^2 \lr[{ \int_0^\infty\dd\chi\  W_g(\chi)~ {^{\tt1}}G_\ell^{\tt -3}(x,\chi)}]\lr[{ \int_0^\infty\dd\chi'\  W_g(\chi')~ {^{\tt1}}G_\ell^{\tt 0}(x,\chi')}]
\end{align}
for a scale-invariant ($n_s=1$) primordial curvature power spectrum. Here, $A = 2\pi^2 A_s $  as previously defined. The quantities in the square brackets are data products generated for our Fisher forecast, with ${^{\tt1}}G_\ell^{\tt 0}(x,\chi)$ the exact case that features a divergent integral. 

\vskip 5pt
To test our numerical convergence, we reconstruct the galaxy power spectrum \eqref{eq:clgg-recon} with the same kernels that we have generated for the Fisher forecast. Throughout this work, we perform the integrals over $x$ and $\chi$ via quadrature with sampling choices outlined as follows. For the $x$ variable, a number of $N_x = 64$ points are sampled over the support of the galaxy window function. To efficiently evaluate \eqref{eq:Ilnt}, we choose to directly sample $t \equiv x /\chi$ with $N_t = N_\chi = 512$ points, whose resolution is adapted to the localisation properties of the kernel $I_L(n,t)$, which is strongly peaked around $t=1$ \cite{Assassi:2017lea}, with its support progressively shrinking with increasing $L$. We follow appendix B of Ref. \cite{Assassi:2017lea}, which quantifies the width of the kernel by introducing $t_{\rm min}(n,L)$, defined by 
\begin{align}
    |I_L(n,t_{\rm min}(n,L))| \equiv 10^{-5}|I_L(n,1)| \;,
\end{align}
and exploit these characteristic scales to construct a piecewise geometric grid over $t\in[0.8,2.2]$. For $t<1$, breakpoints are introduced to gradually increase the grid resolution at the points $t_* \equiv t_{\rm min}(0,L_{\rm jump})$ for the representative multipoles $L_{\rm jump} = 150,1000,3000$, to which we assign correspondingly larger number of sampling points. At $t>1$, we use breakpoints reflected about unity, $2-t_*$. The reflection is a valid approximation to the exact symmetry \eqref{eq:reflect} to leading order in $1-t_*$. This sampling choice  therefore provides high resolution where the high-$L$ kernels vary most rapidly while avoiding unnecessary sampling in regions where the kernels are suppressed. We have assessed that this sampling of the $t$-grid provide good agreement with the Limber-limit behaviour of the kernels in the high-$L$ regime, where the integral over $\chi$ can be done analytically, while remaining numerically converged under further increment in resolution.

\begin{figure}
    \centering
    \includegraphics[width=0.7\linewidth]{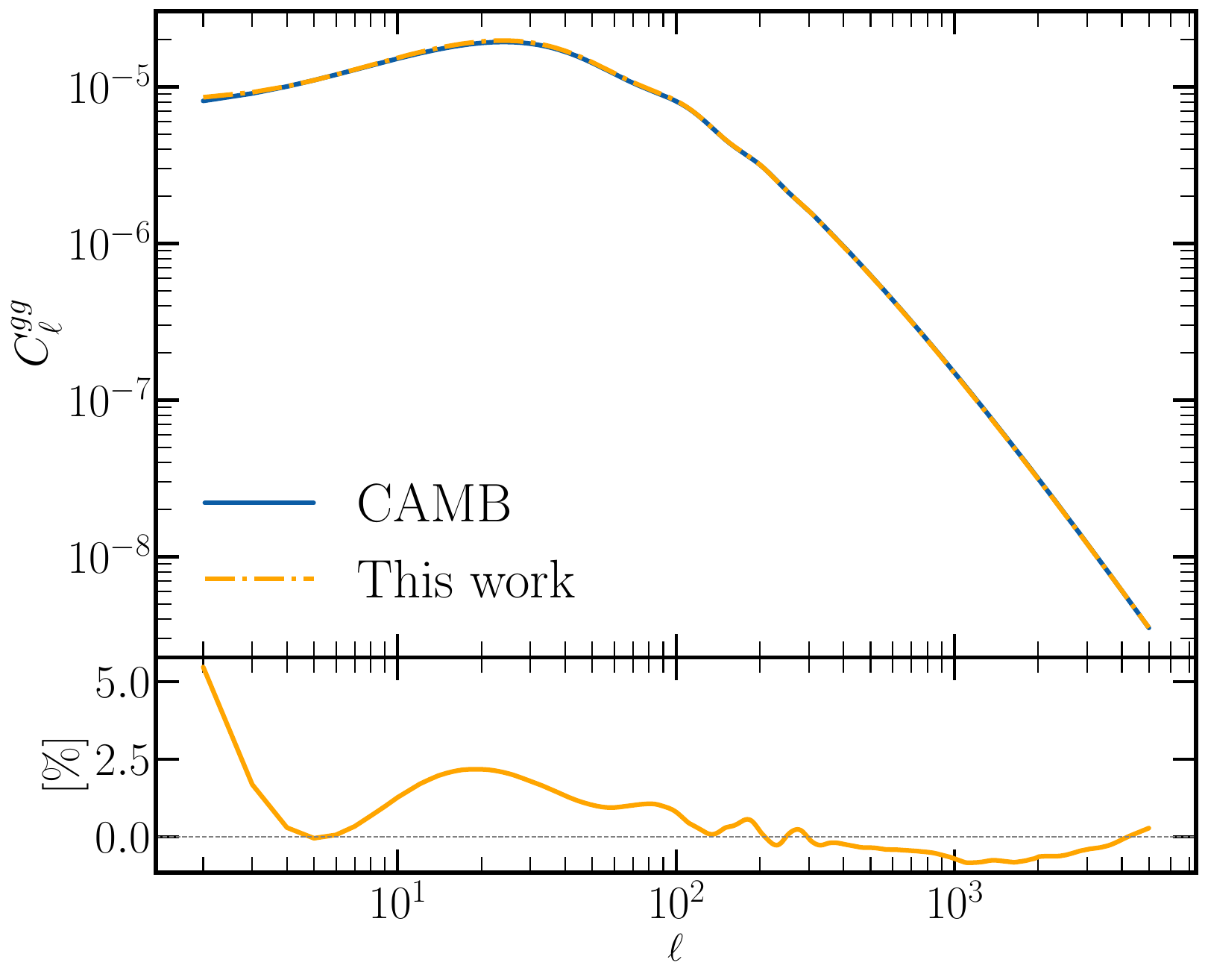}
    \caption{Comparison of galaxy power spectrum reconstructed via \eqref{eq:clgg-recon} to CAMB output. The lower panel shows the percentage difference.}
    \label{fig:cons}
\end{figure}

\vskip 5pt
The galaxy power spectrum reconstructed above can then be compared to that generated via \texttt{CAMB}. The comparison between the two is shown in figure \ref{fig:cons}. Our reconstructed galaxy power spectrum agrees with \texttt{CAMB} to $5\%$ for $\ell=2$, and the deviation is $<2.5\%$ for $\ell\geq 3$. This mild discrepancy primarily arise from (i) a slight imperfection in the FFTLog decomposition near features such as the turnover bend and BAO wiggles, and (ii) a sampling of the $(x,\chi)$-grid outlined above at regions $\chi \gg x$ which is suboptimal for small $\ell$'s where the support of spherical Bessel functions can be very broad. Nevertheless, these deviations are tolerable given the precision of our forecast, and we defer optimisations for future exploration.
% \newpage
%%%%%%%%%%%%%%%%%%%%%%%%%
\phantomsection
%\enlargethispage{\baselineskip}
%\addtocontents{toc}{\protect\enlargethispage{\baselineskip}}
\addcontentsline{toc}{section}{References}
\bibliographystyle{utphys}
{\linespread{1.075}
	\bibliography{Boodds}
}

\end{document}